\documentclass{aastex63}
\usepackage{epstopdf}
\usepackage{natbib}
\usepackage{graphicx}
\usepackage{CJK}

\newcommand       \Teff         {T_{\rm {eff}}}
\defcitealias{2021ApJS..254...38S}{Paper\,I}

\accepted{July 15, 2021}
\submitjournal{ApJS}
\begin{document}
\begin{CJK*}{UTF8}{gbsn}
\title{The Extinction and Distance of the MBM Molecular Clouds at High Galactic Latitude}
\author[0000-0002-2473-9948]{Mingxu Sun (孙明旭)}
\affiliation{Department of Astronomy,
               Beijing Normal University,
               Beijing 100875, China}
\author[0000-0003-3168-2617]{Biwei Jiang(姜碧沩)}
\affiliation{Department of Astronomy,
               Beijing Normal University,
               Beijing 100875, China}
\author[/0000-0003-2645-6869]{He Zhao(赵赫)}
\affiliation{University C\^ote d’Azur, Observatory of the C\^ote d’Azur, CNRS, Lagrange Laboratory, Observatory Bd, CS 34229, 06304 Nice cedex 4, France}

\author[0000-0003-1218-8699]{Yi Ren (任逸)}
\affiliation{Department of Astronomy,
               Beijing Normal University,
               Beijing 100875, China}

\correspondingauthor{B.~W. Jiang}
\email{bjiang@bnu.edu.cn}

\begin{abstract}

Based on the accurate color excess $E_{\rm G_{BP},G_{RP}}$ of more than 4 million stars and $E_{\rm NUV,G_{BP}}$ of more than 1 million stars from \citet{2021ApJS..254...38S}, the distance and the extinction of the molecular clouds in the MBM catalog at $|b|>20^{\circ}$ are studied in combination with the distance measurement of \emph{Gaia}/EDR3. The distance as well as the color excess is determined for 66 molecular clouds. The color excess ratio $E_{\rm G_{BP},G_{RP}}/E_{\rm NUV,G_{BP}}$ is derived for 39 of them, which is obviously larger and implies more small particles at smaller extinction. In addition, the scale height of the dust disk is found to be about 100 pc and becomes large at the anticenter direction due to the disk flaring.
\end{abstract}

\keywords{Distance measure(395); High latitude field(737); Interstellar dust(836); Molecular clouds(1072); Ultraviolet extinction(1738)}

\section{Introduction}

The study of molecular clouds, the site of star formation \citep{1999ASIC..540....3B}, is important for the information of the initial mass function of stars and the build-up of galaxies. Molecular clouds in the Milky way are the nearest and most accessible star forming sites. Carbon monoxide is the main tracer of molecular cloud, which is much easier to be excited and observed than $\rm H_2$, and is used to detect a large number of molecular clouds in many large-scale joint observations of the galaxy \citep[e.g.][]{1985ApJ...295..402M, 1997A&A...327..325M, 2001ApJ...547..792D}. However, the distance as the fundamental and key parameter for studying molecular clouds is often difficult to determine. Many distance measurement methods, such as stellar photometric parallax and period-luminosity relation, are not suitable for molecular clouds.

Previously a popular method is to estimate the distance of clouds by the Galactic kinematics, i.e.  the distance to which  the radial velocity of the cloud corresponds at the rotation curve of the Galactic disk \citep[e.g.][]{1994A&AS..103..541B, 1997A&A...327..325M, 2005PASJ...57..917N, 2009ApJ...699.1153R, 2014ApJS..212....2G, 2017ApJ...834...57M}. This technique is widely applied to estimate the distances to a large number of molecular clouds in the inner disk of the Galaxy \citep[e.g.][]{2009ApJ...699.1153R}. But the well-known problems are the large uncertainty induced by the presence of peculiar and non-circular motions, and the ambiguity that one velocity can correspond to two distances at either side of the tangent point. Another frequently used method is to find the distance to the objects associated with a cloud and to place the cloud at the same distance, for instance, many clouds have produced young OB associations of stars for which distances can be estimated. This method can be applied to some specific cases.

Both of the above two methods are applicable only to the low-latitude clouds in the disk because high-latitude clouds deviate much from the disk rotation curve and own no young massive stars. Since molecular clouds possess not only high-density gas but also high-density dust, their extinction exceeds significantly the surrounding diffuse interstellar medium. Thus the distance to molecular clouds can be inferred from the high extinction caused by them \citep{2009ApJ...692...91G,2017MNRAS.472.3924C}. Early in 1923, \citet{1923AN....219..109W} first effectively describes the Wolf diagrams based on star counts in obscured vs. reference fields to determine the distance to molecular clouds, and \citet{1986A&A...168..271M} applied the Wolf diagrams to a small subset of MBM clouds. According to the two-dimensional (2D) Galactic extinction maps, \citet{2011PASJ...63S...1D} identified  more than 7,000 molecular clouds based on the idea that the high extinction is caused by molecular clouds.  As the extinction map is two-dimensional, no distance information can be obtained.  With the 3D extinction maps which is constructed by comparing the observed color distributions of Galactic giant stars with those predicted by the Galactic model, \citet{2009ApJ...706..727M} catalogued  over 1,000 clouds together with their distance information, the errors of their distances are about 0.5--1 kpc. Comparing the stars in front of the clouds which have little extinction with the predictions of the Galactic model, \citet{2009ApJ...703...52L} and \citet{2011A&A...535A..16L} have estimated the distances of many clouds. With the multi-band photometry by PanSTARRS-1 \citep{2010SPIE.7733E..0EK} and the resultant color indexes of numerous stars,  \citet{2014ApJ...786...29S} derived the distances to 18 well-known star-forming regions, and 108 molecular clouds at high Galactic latitude selected from \citet{1985ApJ...295..402M} and \citet{2001ApJ...547..792D} according to the breakpoint of the extinction.

The distances obtained in these studies are not measured directly but with the help of some stellar or Galactic model and suffer relatively large uncertainty. The \emph{Gaia} mission has changed this situation drastically.  The \emph{Gaia}/DR2 catalog \citep[\emph{Gaia}/DR2;][]{2018A&A...616A...1G} provides  the distances to more than a  billion of stars, which renews the way to determine the distance to molecular clouds by their extinction. With the \emph{Gaia}/DR2 data , \citet{2019ApJ...879..125Z} present a uniform catalog of accurate distances to local molecular clouds according to the breakpoint of the extinction. \citet{2019A&A...624A...6Y} obtain the distances of molecular clouds at high Galactic latitudes ($|b| > 10^{\circ}$) from the parallax \citep{2018A&A...616A...2L} and G-band extinction ($A_G$) measurements of the stars at the cloud's sightline from \emph{Gaia}/DR2. Based on the three-dimensional dust reddening map and estimates of color excesses and distances of over 32 million stars, \citet{2020MNRAS.493..351C} identified 567 dust/molecular clouds which is  within 4 kpc from the Sun at low Galactic latitudes ($|b|\le10^{\circ}$) with a hierarchical structure identification method and obtained their distance estimates by a dust model fitting algorithm. Benefiting from the large number of stars for the individual molecular clouds and the robust estimates of the stellar distances from \emph{Gaia}/DR2, the errors of distances of these works are typically only about 5 per cent.

The high-latitude molecular clouds are generally very optically thin which leads to much smaller extinction than those in the disk, and usually very close, thus a precise measurement of extinction and distance to the stars is necessary to estimate their distance by the extinction method.  Spectroscopy can be used to determine stellar intrinsic color index and thus extinction usually more accurate than multi-band photometry. The general inefficiency of spectroscopy in comparison with photometry is recently compensated by the large-area multi-object spectroscopy such as the LAMOST survey which has observed almost 10 million stars. Using the stellar parameters derived from the LAMOST and GALAH survey,  we (\citet[\citetalias{2021ApJS..254...38S} hereafter]{2021ApJS..254...38S}) determined the color excess $E_{\rm G_{BP},G_{RP}}$  accurate to $\sim0.01$ mag on average towards about four-million stars. In addition, the color excess $E_{\rm NUV,G_{BP}}$ is calculated for more than one million stars accurate to $\sim0.1$ mag. Moreover, the \emph{Gaia}/EDR3 is recently released with apparently more precise measurement of stellar parallaxes and distances. The combination of the accurate color excess and distance brings about the possibility to determine the distances to the molecular clouds at high-latitude.  In addition, the dust property in the high-latitude clouds may be inferred from the color excess ratio of $E_{\rm G_{BP},G_{RP}}/E_{\rm NUV,G_{BP}}$.  \citet{1992ApJ...393..193W} study the low-resolution UV spectra of the B3 V star HD 210121 located behind the high-latitude molecular cloud DBB 80 and  yield an extinction curve with the very steep rise in far UV. The extremely steep far-UV extinction and the augment of the intensity at 12 $\mu$m are consistent with the presence of an enhanced population of very small grains.  We (\citetalias{2021ApJS..254...38S})  also find that there may be more small dust grains at high than at low galactic latitude, which is supported by the steeper increase of extinction towards the FUV band.

This paper is part of an ongoing project to study the extinction and dust as well as the 3D distribution of molecular clouds at high latitude based on the spectroscopic and astrometric measurements of stars. This work focuses on the MBM high-latitude molecular clouds.

\section{Sample and Data}\label{DATA}

\citet{1984ApJ...282L...9B} started a project to conduct a systematic search of high-latitude molecular clouds (MCs) by the CO line observation of potentially obscured regions identified from the POSS prints, which is followed and completed by   \citet{1985ApJ...295..402M} (MBM)\footnote{\url{http://simbad.u-strasbg.fr/simbad/sim-ref?querymethod=bib&simbo=on&submit=submit+bibcode&bibcode=1985ApJ...295..402M}}. This project resulted in 124 detections of MCs with $|b|>20^{\circ}$. We adopt the center positions of the MCs from MBM, and the size as well if given. For the 88 MCs whose size is unavailable by MBM, a radius of $90'$ is taken as default, which is approximate to the average size of the high-latitude MCs \citep{2002A&A...383..631D}.

The tracers for the extinction and distance of the high-latitude clouds are chosen from \citetalias{2021ApJS..254...38S}.  Using the blue-edge method (c.f. \citet{2017AJ....153....5J} and \citet{2018ApJ...861..153S}), \citetalias{2021ApJS..254...38S} calculated the color excess with respect to the \emph{Gaia}/$\rm G_{BP},G_{RP}$ and \emph{GALEX}/NUV bands, i.e. $E_{{\rm G_{BP},G_{RP}}}$ and $E_{{\rm NUV,G_{BP}}}$ of more than 4 million and 1 million dwarfs respectively, which are mostly located at high-latitude. These color excesses are determined from the intrinsic and observed color indexes, in which the intrinsic color index is  determined from the stellar parameters $\Teff$ , $\log g$ and metal abundance $Z$ from the LAMOST/DR7 and GALAH/DR3 spectroscopic surveys. The average error of $E_{{\rm G_{BP},G_{RP}}}$ and $E_{{\rm NUV,G_{BP}}}$ is $\sim$ 0.01 mag and 0.1 mag respectively. Besides, the  distances of those sources are obtained  by the corresponding parallax in  \emph{Gaia}/EDR3 \citep{2020arXiv201201533G}.

\begin{figure}
\centering
\centerline{\includegraphics[scale=1]{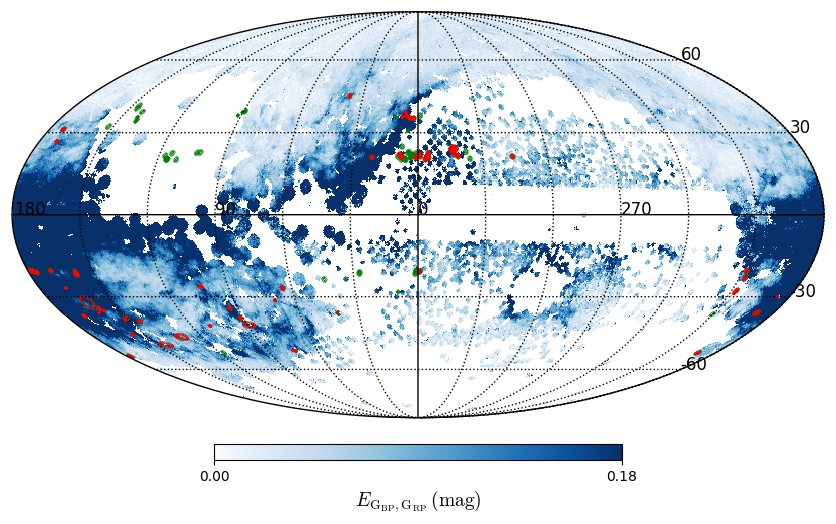}}
\caption{Distribution of the MBM molecular clouds (ellipses) in the extinction map (blue background) expressed by  $E_{\rm G_{BP},G_{RP}}$ from \citetalias{2021ApJS..254...38S}, where red and green represents the molecular clouds within and outside the extinction map respectively.
\label{fig1}}
\end{figure}

The cross-match between the MBM molecular clouds and the stars in \citetalias{2021ApJS..254...38S} finds that there are 75 molecular clouds to whose sightline $E_{{\rm G_{BP},G_{RP}}}$ of 13,773 stars is available,  and 47 molecular clouds of them to whose sightline   $E_{{\rm NUV,G_{BP}}}$ of 3,614 stars is available as well. The distribution of all the $E_{{\rm G_{BP},G_{RP}}}$ in \citetalias{2021ApJS..254...38S} and MBM molecular clouds are displayed in Figure~\ref{fig1}, where red and green ellipses represent the MCs within and outside the extinction map respectively.

\section{method}\label{method}

\subsection{Selection of the reference and cloud region}

\subsubsection{The Planck 857GHz map}\label{extinction and emission}

The extinction of the cloud is the difference of the post-cloud  from the pre-cloud extinction. Then it is necessary to delineate the region of the cloud and the background for reference in order to determine the extinction caused by the MC. The area given by MBM makes a very good initial value for the region of the cloud, but no reference region is available. Though the region in front of the cloud can be taken for the reference as \citet{2020ApJ...891..137Z} and \citet{2018ApJ...855...12Z} have done, the high-latitude MCs are usually close and the small-range foreground stars can hardly reflect the global trend of the extinction variation along the sightline with no cloud. Thus an independent region other than the cloud region is selected as the reference. Moreover, the cloud normally has irregular shape that an ellipse cannot describe the true boundary of the cloud. We determine a precise border of the cloud according to the infrared emission image instead of the molecular emission because infrared emission comes directly from dust and is proportional to extinction.

The Planck 857 GHz image \citep{2020A&A...641A...1P}, closely correlated with the dust emission, is used to mark the reference and molecular cloud region. The Planck 857 GHz survey has a similar spatial resolution ($5'$) as  the IRAS 100 $\mu$m image \citep{1998ApJ...500..525S,2005ApJS..157..302M}, while its sensitivity is much higher. In comparison with the CO survey \citep{2001ApJ...547..792D}, the Planck 857 GHz survey is more complete at high Galactic latitude.

\begin{figure}
\centering
\centerline{\includegraphics[scale=0.8]{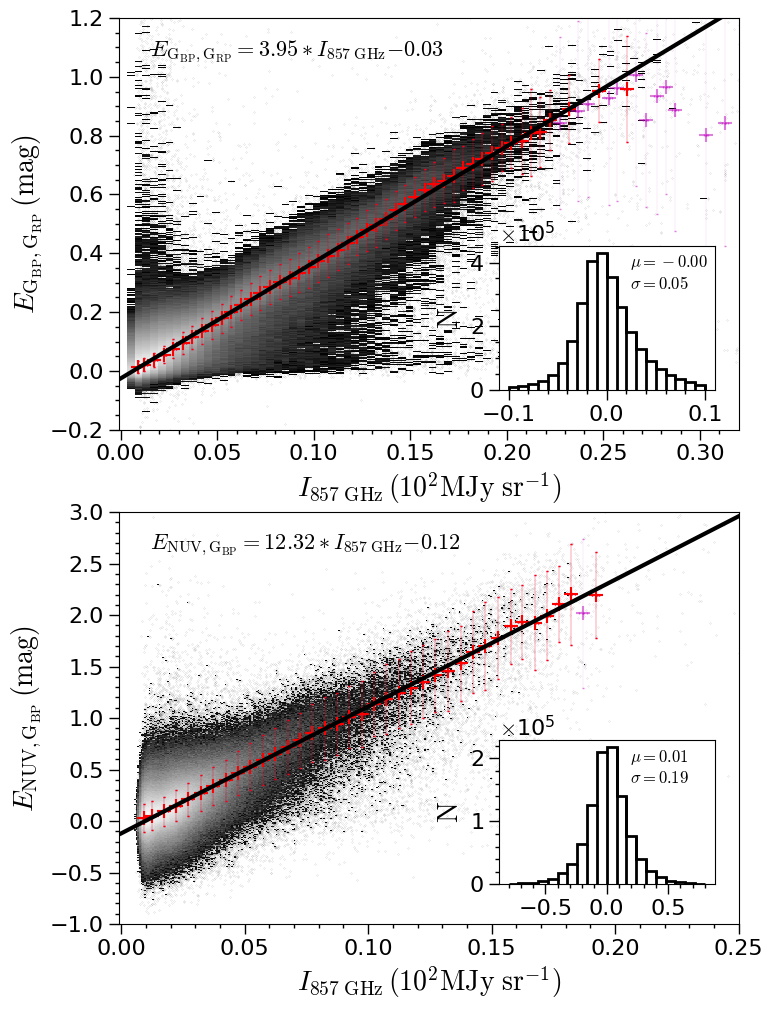}}
\caption{Linear fitting of the color excess $E_{{\rm G_{BP},G_{RP}}}$ to the Planck/857GHz intensity $I_{857 \rm GHz}\,(10^2{\rm MJy\ sr^{-1}})$ (top) and $ E_{{\rm NUV,G_{BP}}}$ to $I_{857 \rm GHz}\,(10^2{\rm MJy\ sr^{-1}})$ (bottom). The gray-scale decodes the source density, the red and magenta crosses denote the median values of each bin with small and large deviations from the linear fitting line (see \ref{extinction and emission} for details). The inset shows the distribution of the residuals with its median and standard deviation.
\label{fig2}}
\end{figure}

In order to see the relation between the 875 GHz (350 $\mu$m) cumulative dust emission from \citet{2020A&A...641A...1P} and the color excess,  the sources from \citetalias{2021ApJS..254...38S} are selected only when they have latitude $|b| >20^{\circ}$ and Galactic plane distance $|h|>200$ pc so that their extinction can be considered as cumulative along the specific sightlines. The comparison of the extinction with the dust emission, as shown in Figure~\ref{fig2}, finds a very tight linear relationship between them. The quantitative relation is yielded in the same way as in \citetalias{2021ApJS..254...38S} by iteratively clipping stars beyond 3$\sigma$ of the median, which results in  $ E_{\rm G_{BP},G_{RP}}/I_{857 \rm GHz}\,(10^2{\rm MJy sr^{-1}})= 3.95$, and $E_{\rm NUV,G_{BP}}/I_{857 \rm GHz}\,(10^2{\rm MJy sr^{-1}})=$ 12.32.

\subsubsection{Selection of the reference and cloud region}

\begin{figure}
\centering
\centerline{\includegraphics[scale=0.25]{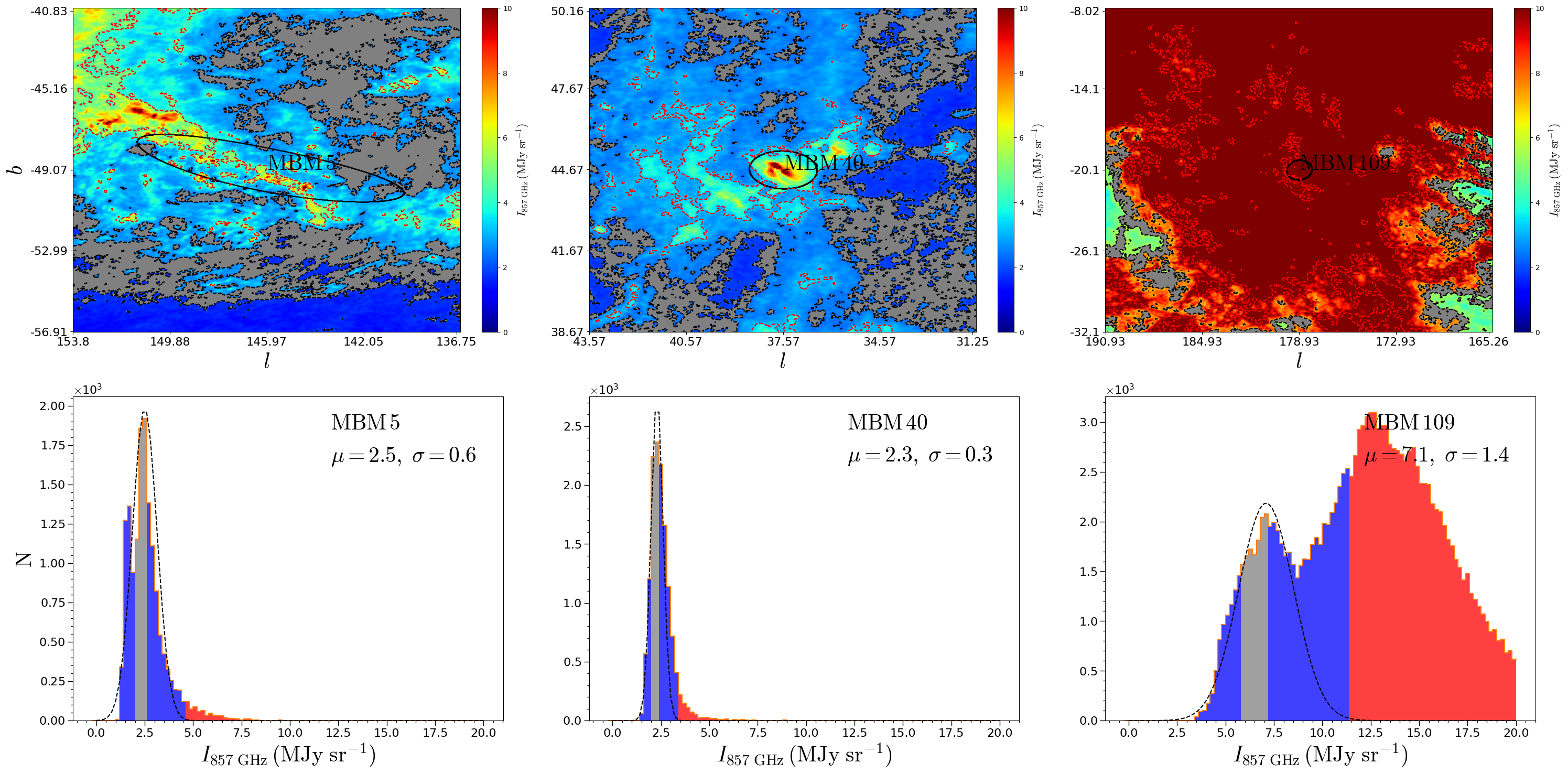}}
\caption{The studied area (top) and histogram (bottom) of $I_{857 \rm GHz}\,({\rm MJy\ sr^{-1}})$ in MBM5 (left), MBM40 (middle) and MBM109 (right). In the studied area, the  gray filled region bordered by black dashed line is the $background\,\, region$, and the green-yellow-red region bordered by red dashed line region inside the MBM cloud region (black solid line) is the $cloud\,\, region$.  In the histogram, the black dashed curve is the local gaussian fit, where the blue bars represent the noise (left) and transition (right) region, the gray bars represent the $background\,\, region$ sources and the red bars represent the $cloud\,\, region$ sources. 
\label{fig3}}
\end{figure}

The denser MCs are supposed to have higher infrared intensity than the reference region so that the areas can be determined according to the Planck/875 GHz intensity. At first,  both the reference and cloud region is searched in a square area with  a side length of 4 times the cloud's equivalent angular diameter ($\sqrt{\theta_{\rm major} \times \theta_{\rm minor}}$) centering at the cloud position. However, $I_{857 \rm GHz}$ does not decrease significantly within this area in many cases. So the region is expanded to $m$ times of the cloud's angular diameter until an appropriate region can be found for reference. The technical route of selecting the areas is illustrated by taking three typical cases (MBM5, MBM109 and MBM40) as examples in Figure~\ref{fig3}.

The distribution of the small peak is fitted by a Gaussian function to determine the median ($\mu$) and the standard deviation ($\sigma$) of $I_{857 \rm GHz}$ for the reference region; then the region with $I_{857 \rm GHz}$  in the range of $\mu-\sigma$ to $\mu$ is selected as the $reference\,\, region$, denoted by the black dashed line and gray histogram in the upper and lower panel respectively of Figure~\ref{fig3}, and the region with $I_{857 \rm GHz}$ above $\mu+3\sigma$ (the red dashed line and histogram in the upper and lower panel respectively of Figure~\ref{fig3}) and inside the MBM-assigned MC area (marked by the black solid-line ellipse) is selected as the $cloud\,\, region$.

The MBM5 is a relatively isolated object for which the reference region can be found in an area as four times big as the cloud area. But MBM 109 is located within a large high-$I_{857 \rm GHz}$ area towards the sightline of the Tau-Per-Aur complex cloud so that the reference area has to be found in a far area, specifically, the area is expanded to 16 times the cloud's angular diameter. MBM40 is taken as an example because it will be shown later that its distance is only 63 pc, which might put it inside the Local Bubble or right at the boundary of the Local Bubble. It can be seen that the values of $\mu$ and $\sigma$ depend on the property and location of MC. In this way, the reference and cloud area is defined for each cloud in an area specified by the $m$ value listed in Table~\ref{table1}.

\subsection{The Extinction-jump Model}

Under the assumption that interstellar extinction increases smoothly with distance in the absence of any molecular clouds, the extinction will make an upward jump at the distance of the cloud with the presence of a molecular cloud. In order to obtain accurate distance and extinction of molecular clouds, we take  the extinction-jump model in \citet{2020ApJ...891..137Z}, which is insensitive to the outliers. This model was designed originally by \citet{2018ApJ...855...12Z} and improved by \citet{2020ApJ...891..137Z} and used to determine the distance and the extinction  of Galactic supernova remnants (SNRs). Similar model is used by \citet{2017MNRAS.472.3924C} and \citet{2019MNRAS.488.3129Y} for other supernova remnants as well as molecular clouds. The MCs cause the same effect as SNRs on the  extinction and thus the extinction-jump model is applicable. In details,  the total extinction in terms of color excess $E (d)$ towards the sightline of the molecular cloud is composed of two parts: the color excess of the cloud $E^{\rm MC}(d)$ which dominates the total extinction and the color excess of the diffuse interstellar medium $E^{\rm DISM}(d)$ as follows:
\begin{equation}\label{JUMP1}%
E(d)= E^{\rm DISM}(d)+E^{\rm MC}(d).
\end{equation}
Moreover, $E^{\rm MC}$(d) is described by an erf function,
\begin{equation}\label{JUMP2}%
E^{\rm MC}(d)=\Delta E^{\rm MC} \times [1+erf( \frac{d-d_c}{\sqrt{2}\delta d})]
\end{equation}
where $\Delta E^{\rm MC}$ is the amplitude of the color excess jump, i.e. $\Delta E^{\rm MC}_{\rm G_{BP},G_{RP}}$ or $\Delta E^{\rm MC}_{\rm NUV,G_{BP}}$ caused by the cloud,  $d_c$ is the distance to the center and $\delta d$ is  the radius of the cloud calculated from  $d_c \times \theta_c$ with $\theta_c$ being the cloud's angular diameter.

However, unlike \citet{2017MNRAS.472.3924C}, \citet{2019MNRAS.488.3129Y} and \citet{2020ApJ...891..137Z} which use a two-order polynomial function or root function, we use an exponential law  to fit the color excess caused by the diffuse interstellar dust,
 \begin{equation}\label{JUMP3}%
E^{\rm DISM}(d)=E^{\rm 0} \times(1-e^ {\frac{-d}{h'}})
\end{equation}
The function is modified to be in line with the high-latitude location of the MCs to which sightline the material (including the interstellar dust) density falls exponentially with the vertical distance from the Galactic plane. Under this form, the parameter $E^{\rm 0}$ reflects the cumulative color excess, and $h'\times sin(b)$ with $b$ being the latitude of the cloud  is the scale height ($h$) of the dust disk in the sightline.

 \subsection{The model fitting}\label{routine}

The model fitting is performed by the Markov Chain Monte Carlo (MCMC) procedure \citep{2013PASP..125..306F}. In order to set the initial parameters highly reasonable,  a Markov chain is first run with 100 walkers and 250 steps. Then, the final chain uses the initial parameters with 100 walkers and 2000 steps and we choose the last 1750 steps from each wallker to sample the final posterior. The best estimates are the median values (50th percentile) of the posterior distribution and the  uncertainties are derived from the 16th and 84th percentile values. 

The variation of stellar extinction with distance is fitted separately for the  $reference\,\, region$ and the $cloud\,\, region$. Because the sample becomes more and more incomplete with distance, only the objects closer than 2000 pc with relative distance uncertainty  $<30\%$ are taken into account.  The fitting steps are in the following order: (1) Fitting the optical extinction - distance measurements, i.e  the $E_{\rm G_{BP},G_{RP}}$ vs. $d$ points of the reference region by Eq. \ref{JUMP3} to obtain the parameters $E_{\rm G_{BP},G_{RP}}^{\rm 0}$ and $h'_{\rm G_{BP},G_{RP}}$ to describe the change of $E_{\rm G_{BP},G_{RP}}$ with $d$ in the reference region; (2)  Fitting the optical extinction - distance measurements of the cloud region by Eq. \ref{JUMP1} and \ref{JUMP2} after substituting the above values of $E_{\rm G_{BP},G_{RP}}^{\rm 0}$ and $h'_{\rm G_{BP},G_{RP}}$ into  Eq. \ref{JUMP1}, which yield the distance $d_c$ and the optical jump $\Delta E^{\rm MC}_{\rm G_{BP},G_{RP}}$ of the molecular cloud; (3) The same as Step (1) but replacing $E_{\rm G_{BP},G_{RP}}$ by $E_{\rm NUV,G_{BP}}$ of the reference region sources to obtain $E_{\rm NUV,G_{BP}}^{\rm 0}$ and $h'_{\rm NUV,G_{BP}}$; and (4) Similar to Step (2), but replacing the optical parameters by ultraviolet values, i.e. $E_{\rm G_{BP},G_{RP}}$ by $E_{\rm NUV,G_{BP}}$ and $E_{\rm G_{BP},G_{RP}}^{\rm 0}$ and $h'_{\rm G_{BP},G_{RP}}$  by $E_{\rm NUV,G_{BP}}^{\rm 0}$ and $h'_{\rm NUV,G_{BP}}$. In addition, the value of $d_c$  resultant from optical color excess is adopted other than fitted because the number of the measurements in the UV band is only about a quarter in the visual, thus only the jump $ \Delta E^{\rm MC}_{\rm NUV, G_{BP}}$ is derived in this step.

The model fitting for MBM5, MBM40 and MBM109 with the reference and cloud sources are displayed as the example  in Figure~\ref{fig4} and Figure~\ref{fig5} for $\Delta E^{\rm MC}_{{\rm G_{BP},G_{RP}}}$ and $\Delta E^{\rm MC}_{\rm NUV,G_{BP}}$ respectively, where the green dots denote the sources in the reference region used to determine the variation of extinction with distance in the diffuse medium, and the red dots denote the sources in the cloud region used to determine the distance and extinction of the cloud. The key parameters with the uncertainty derived from modelling are shown at the upper left corner of the figures (An extended version of Figure 4 is available online for all the sample clouds).

\section{Results and Discussions}

\subsection{The distance}

\begin{figure}
\centering
\centerline{\includegraphics[scale=0.75]{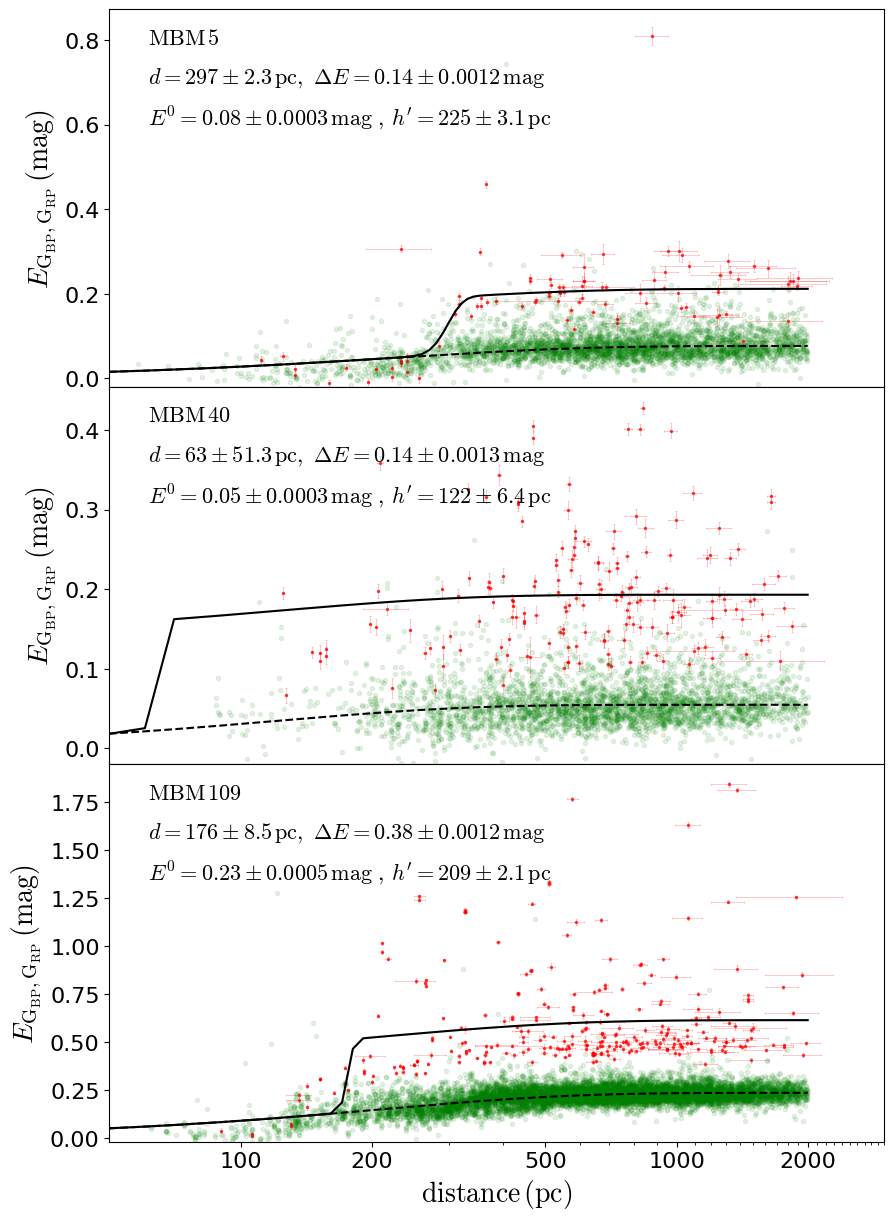}}
\caption{The fitting to the color excess, $E_{\rm G_{BP},G_{RP}}$, variation with distance of the stars in the reference (green dots) and the cloud (red dots) region for the three typical clouds ---  MBM5, MBM40 and MBM109, $E_{\rm G_{BP},G_{RP}}$ by the extinction$-$distance model (Eqs.\ref{JUMP1}, \ref{JUMP2} and \ref{JUMP3}).  The parameters derived are shown in the upper left corner, where `d' is the distance, `$\Delta$E' is the color excess jump ($\Delta E^{\rm MC}_{\rm G_{BP},G_{RP}}$), $E^{\rm 0}$ and $h'$ are the foreground parameters of the molecular cloud ($E_{\rm G_{BP},G_{RP}}^{\rm 0}$ and $h'_{\rm G_{BP},G_{RP}}$). (An extended version of this figure for all the studied clouds is available on line.)
\label{fig4}}
\end{figure}

\begin{figure}
\centering
\centerline{\includegraphics[scale=0.75]{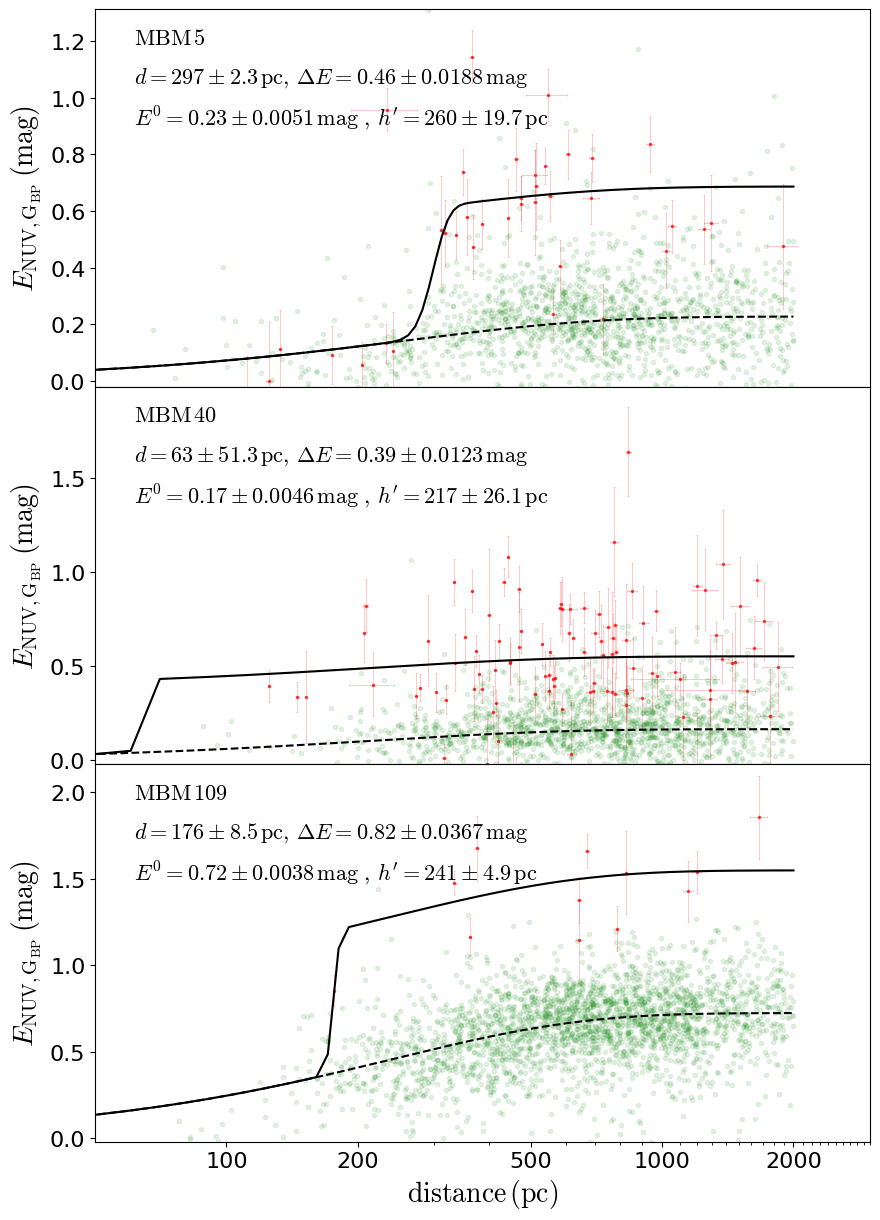}}
\caption{The same as Figure~\ref{fig4}, but for $E_{\rm NUV,G_{BP}}$.
\label{fig5}}
\end{figure}

\begin{figure}
\centering
\centerline{\includegraphics[scale=1]{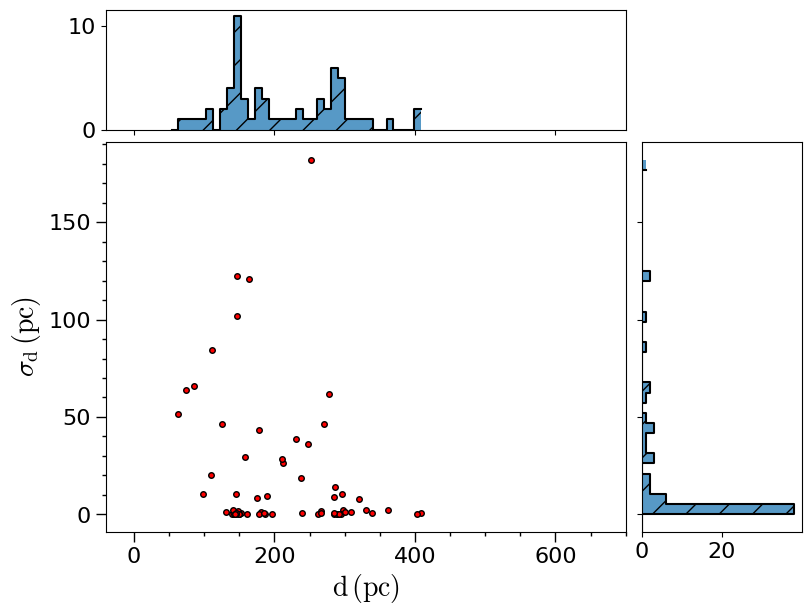}}
\caption{The distribution of the distances (d) and their uncertainties ($\sigma_d$) .
\label{fig6}}
\end{figure}

The distance is derived for 66 among the 75 MBM clouds towards which sightline more than three stars are present with the optical color-excess measurement and lying behind the molecular cloud. The derived parameters are tabulated in Table~\ref{table1}. The distribution of the distances and their uncertainties are shown in Figure~\ref{fig6} where the symbols obey the convention in Figure~\ref{fig4}.  A simple visual inspection would conclude that the fitting agrees very well with the measurements in most cases. Meanwhile the distances to G37.57-35.06 (MBM11), G157.98+44.67 (MBM40), G341.61+21.40 (MBM119), G343.28+22.12 (MBM123) and G359.48-20.47 (MBM152) seem to be under-estimated because there are few sources at the very close distance. As mentioned earlier, MBM40 is the nearest with a  distance of only $63\pm51$ pc, which put is inside the Local Bubble or right at the boundary of the Local Bubble if being true. Though the uncertainty is large, the best value is consistent with \citet{2014ApJ...786...29S}. On the other hand, this value is smaller than 93 pc yielded by \citet{2019ApJ...879..125Z}. Indeed, Figure \ref{fig4} and \ref{fig5} indicate that the first stars that show a marked increase in color excess have a distance of about {125 pc}, apparently much larger, however, still within the range of the uncertainty. These cases indicate that a precise distance by our method needs a continuous distribution in distance of the tracers in particular around the jump point. For MBM 40, more objects in front of the cloud should be measured to confirm its close distance.

\begin{figure}
\centering
\centerline{\includegraphics[scale=0.85]{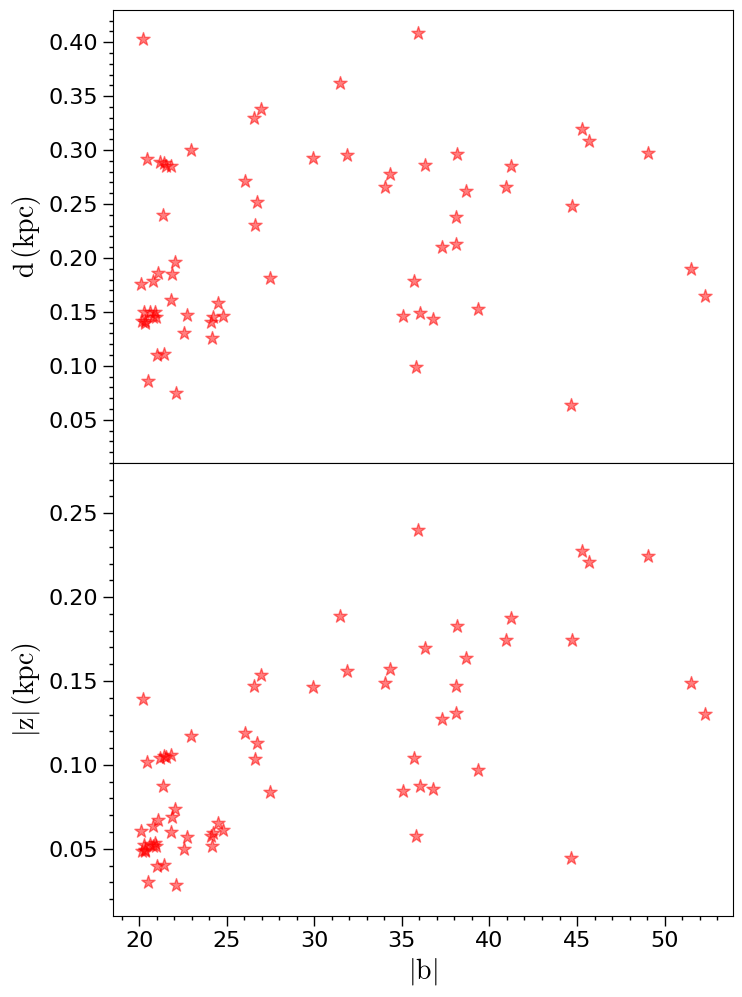}}
\caption{The distance d (upper panel) and the Galactic disk distance $\rm |z|$ (lower panel) versus the Galactic latitude $|b|$.
\label{fig7}}
\end{figure}

Figure~\ref{fig7} shows the distance $d$ and the vertical distance to the Galactic plane $|z|$ versus the Galactic latitude $|b|$. There is no systematic trend of $d$ with $|b|$ expected from that these clouds are local, while $|z|$ increases with the Galactic latitude $|b|$.

\begin{figure}
\centerline{\includegraphics[scale=1]{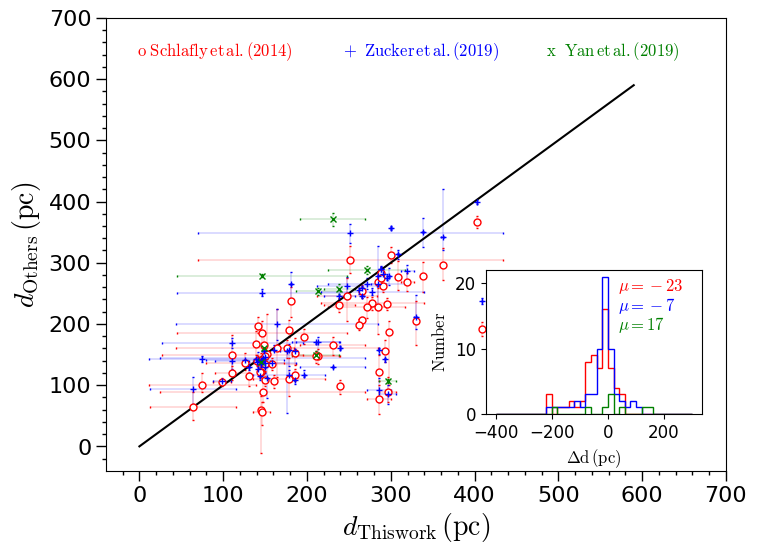}}
\caption{The comparison of the distances of molecular clouds with those obtained by \citet{2014ApJ...786...29S}, \citet{2019A&A...624A...6Y} and \citet{2019ApJ...879..125Z}. The inset is the distribution of the difference with their results.
\label{fig8}}
\end{figure}

The distances to nine of the 66 molecular clouds were measured  by \citet{2019A&A...624A...6Y}, and 64 of them by \citet{2014ApJ...786...29S} and \citet{2019ApJ...879..125Z},  which is compared with ours in Figure~\ref{fig8}. It can be seen that the distances are more or less identical between the works at $d < 200$pc. When $d > 200$pc, this work yields systematically larger distance than the others by different extent in that the difference with \citet{2014ApJ...786...29S} is the largest and with the other two works are mostly within the uncertainties.  Comparing with these works, this work differs in a few aspects: (1) the intrinsic color indexes are derived from spectroscopy other than photometry, (2) the stellar distance comes from the \emph{Gaia}/EDR3 catalog instead of the DR2 catalog, (3) The model considers the thickness of the molecular cloud, and  (4) the rise of the color excess of the foreground sources with distance is considered separately which prevents the premature occurrence of jumps in some molecular clouds. The first two factors improve the accuracy but should have no systematic influence on the distance. 

\subsection{The extinction}

\subsubsection{The cloud}

\begin{figure}
\centering
\centerline{\includegraphics[scale=0.9]{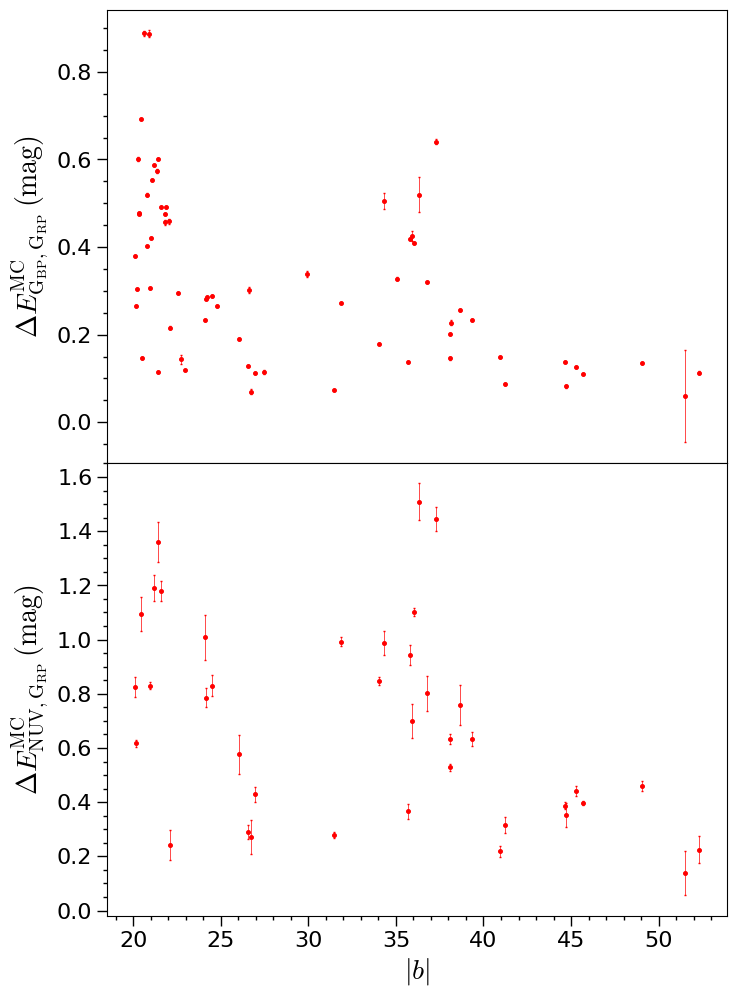}}
\caption{The change of $\Delta E^{\rm MC}_{\rm G_{BP},G_{RP}}$ and $\Delta E^{\rm MC}_{\rm NUV,G_{BP}}$ with the Galactic latitude $|b|$.
\label{fig9}}
\end{figure}

The extinction is determined for 66 MCs expressed by the optical color excess $\Delta E^{\rm MC}_{\rm G_{BP},G_{RP}}$  and for 39 MCs by the UV-optical color excess $\Delta E^{\rm MC}_{\rm NUV,G_{BP}}$. Their distribution along the latitude is shown in Figure~\ref{fig9} that the increase at low latitudes is visible as expected from their smaller vertical distance to the Galactic plane evidenced in the lower panel of Figure \ref{fig7}. Meanwhile, it should be noted that the extinction appears large around $|b| \sim 35^{\circ}-40^{\circ}$. The majority of $\Delta E^{\rm MC}_{\rm G_{BP},G_{RP}}$ is not bigger than 0.3 mag, i.e $E_{\rm G_{BP},G_{RP}}^{\rm MC} \sim $ 1 mag, and some clouds cause a large extinction but still smaller than $E_{\rm G_{BP},G_{RP}}^{\rm MC} \sim $ 3mag. Therefore, most of the clouds may be classified as translucent, and some of them as diffuse molecular clouds.

\begin{figure}
\centering
\centerline{\includegraphics[scale=0.85]{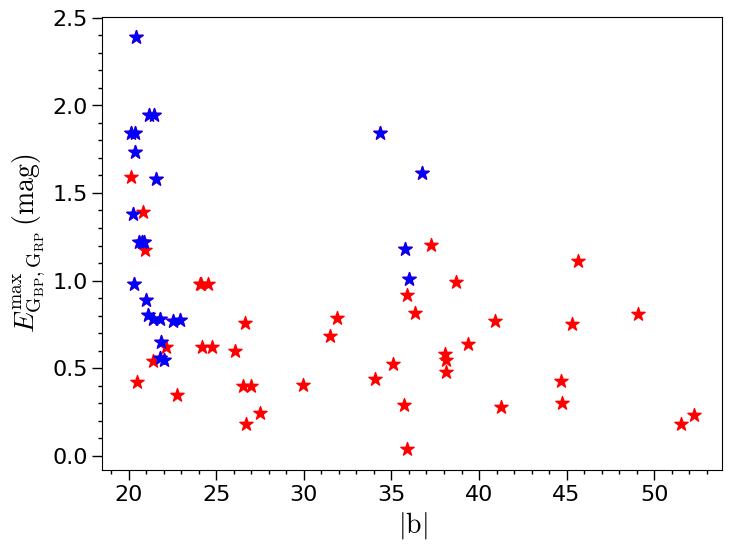}}
\caption{The change of maximum extinction $E_{\rm G_{BP},G_{RP}}^{\rm max}$ in the sightline of molecular clouds with the Galactic latitude $\rm |b|$. The blue and red asterisks are the result of MBM clouds  with or without Lynds dark cloud.
\label{fig10}}
\end{figure}

It should be noted that there are some stars with large color excess which is not reflected in the fitting parameters. The cross-identification with the Lynds dark clouds \citep{1962ApJS....7....1L} finds that 24 MBM clouds contain 20 Lynds dark clouds, which are listed in Table~\ref{table2}. Some MBM clouds cover multiple Lynds clouds while some Lynds clouds repetitively appear in various MBM clouds. Such confusion implies that the boundary of a cloud needs to be re-defined, and will be considered in our next work.  Figure~\ref{fig10} shows the maximum stellar $E_{\rm G_{BP},G_{RP}}$, i.e. $ E^{\rm max}_{\rm G_{BP},G_{RP}}$  behind each MBM cloud identifiable in the Lynds dark clouds catalog in comparison with all the other clouds, where the blue asterisks denote the MBM molecular cloud containing some Lynds dark cloud(s). It can be seen that the majority of these clouds have $E^{\rm max}_{\rm G_{BP},G_{RP}}> 1.0$, i.e. $A_{\rm V}> 2.0$mag. Since dark clouds are normally defined to have $A_{\rm V} > 5$ mag, this is not consistent with the expectation for a dark cloud. It is likely that an interstellar cloud might have an average extinction of 2 mag with small patches having extinctions of 5 mag or more and so, on the basis of these small patches, the cloud is defined as a dark cloud while its average extinction is more like that of a translucent cloud. Meanwhile, a few clouds have $A_{\rm V}\sim 1.0-2.0$mag, smaller than the extinction that a dark cloud should have. One possible reason is that the stars that suffer serious extinction may become too faint to be observable by the LAMOST or GALAH spectroscopy survey. This also manifests that the method of this work derives the median rather than the highest extinction of cloud, so that it is more appropriate for relatively large clouds than smaller dense clouds. Additionally, several MBM clouds have no associated Lynds dark clouds but with $E^{\rm max}_{\rm G_{BP},G_{RP}}> 1.0$, still an acceptable extinction ($A_{\rm V}< 3.0$ mag) for a translucent cloud though we cannot exclude the possibility of the presence of some dark nebula.

\subsubsection{The reference region}

The reference region is selected from the low-intensity noise-like emission at Planck/857GHz, representative of the diffuse interstellar medium. Our model of the reference extinction by Eq.~\ref{JUMP3} assumes an exponential disk with the vertical height. The parameter $E_0$ in Eq.~\ref{JUMP3} is the cumulative extinction and $h'\sin(b)$ is the scale height of the extinction/dust disk towards the specific high-latitude  sightline.

\begin{figure}
\centering
\centerline{\includegraphics[scale=0.5]{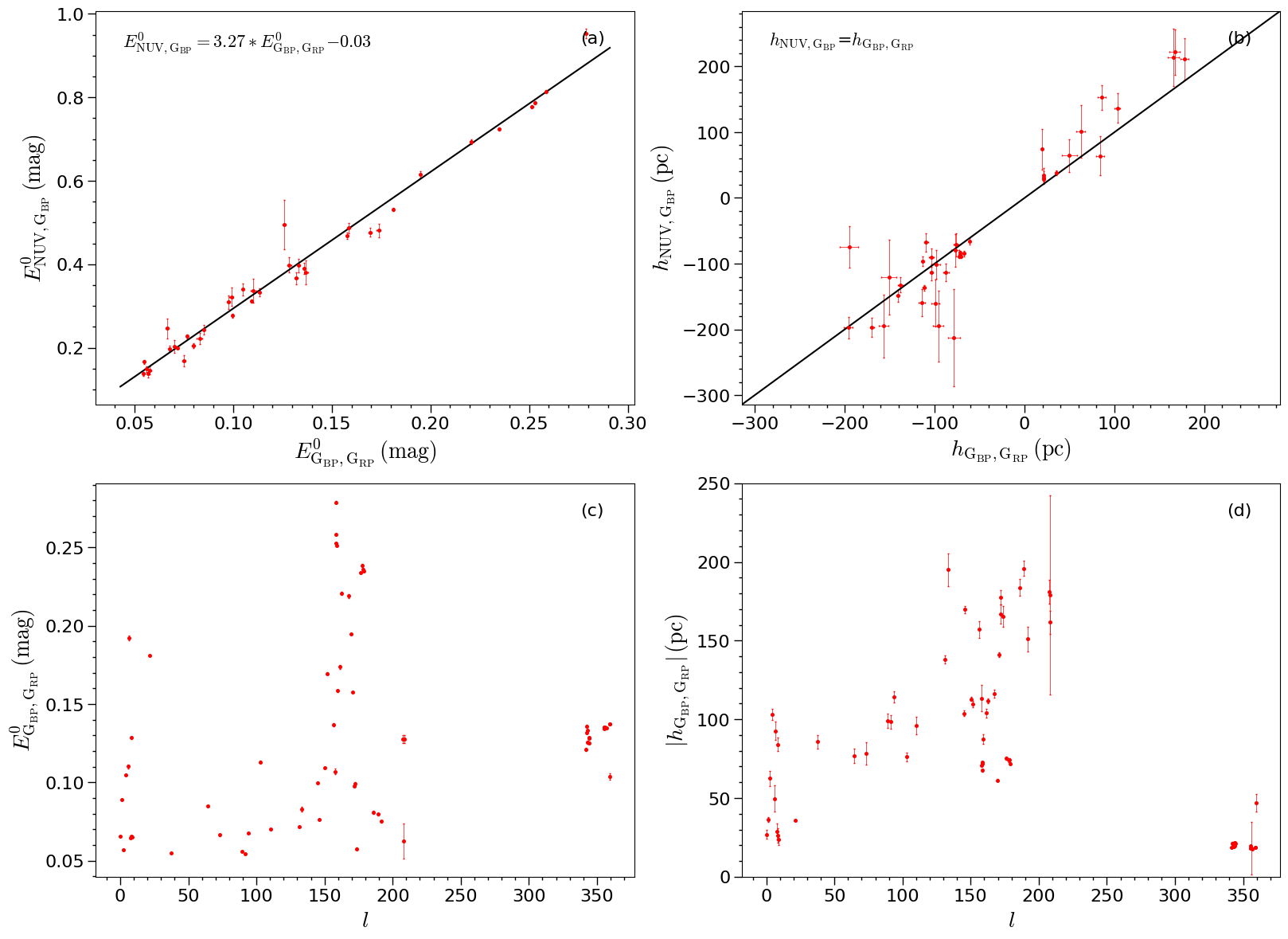}}
\caption{The relationship of $E_{\rm NUV,G_{BP}}^{\rm 0}$ and $E_{\rm G_{BP},G_{RP}}^{\rm 0}$ (top, left) and $h_{\rm NUV,G_{BP}}$ and $h_{\rm G_{BP},G_{RP}}$ (top, right) as well as the distribution of $E_{\rm G_{BP},G_{RP}}^{\rm 0}$(bottom, left) and $|h_{\rm G_{BP},G_{RP}}|$(bottom, right). Red asterisks are the parameters of foreground and black line is the fitting line of $E_{\rm NUV,G_{BP}}^{\rm 0}$ and $E_{\rm G_{BP},G_{RP}}^{\rm 0}$ (top, left)  and $h_{\rm NUV,G_{BP}}$=$h_{\rm G_{BP},G_{RP}}$ line (top, right).
\label{fig11}}
\end{figure}

The relationship between the parameters from fitting $E_{\rm G_{BP},G_{RP}}^{\rm 0}$ and $E_{\rm NUV,G_{BP}}^{\rm 0}$ for the reference regions is shown in Figure~\ref{fig11}. As shown in Figure~\ref{fig11} (a), there is a good linear relationship between $E_{\rm G_{BP},G_{RP}}^{\rm 0}$ and $E_{\rm NUV,G_{BP}}^{\rm 0}$. The slope of the fitting line is 3.27, which reflects the ratio of the accumulated color excess of the diffuse interstellar medium in the UV and optical band. This ratio is very close to the all-sky color excess ratio of 3.25 by \citetalias{2021ApJS..254...38S}. 

After multiplying $h'_{\rm G_{BP},G_{RP}}$ and $h'_{\rm NUV,G_{BP}}$ by $\sin(b)$, the dust disk scale hight $h_{\rm G_{BP},G_{RP}}$ and $h_{\rm NUV,G_{BP}}$ in the sightline is obtained. Figure~\ref{fig11} (b) presents the relation between $h_{\rm G_{BP},G_{RP}}$ and $h_{\rm NUV,G_{BP}}$, which are basically distributed near the equal line while some points deviate. Because the parameter $h'$ is very sensitive to the data size,  $h_{\rm G_{BP},G_{RP}}$ should be more reliable due to its much smaller error and more data points in the \emph{Gaia}/EDR3 catalog.  The change of $E_{\rm G_{BP},G_{RP}}^{\rm 0}$ and $|h_{\rm G_{BP},G_{RP}}|$ with Galactic longitude is displayed in Figure~\ref{fig11} (c) and Figure~\ref{fig11} (d). The scale height varies from about 50 pc to 250 pc, with an obvious increase towards the anticenter direction consistent with a flaring dust disk. This median thickness of about 100 pc indicates that the dust disk agrees with the thin gaseous disk of the Milky Way whose scale height ranges around 100-300 pc \citep[e.g.][]{2017ApJ...843..141F,2017MNRAS.467.2430M}.

\begin{figure}
\centering
\centerline{\includegraphics[scale=1]{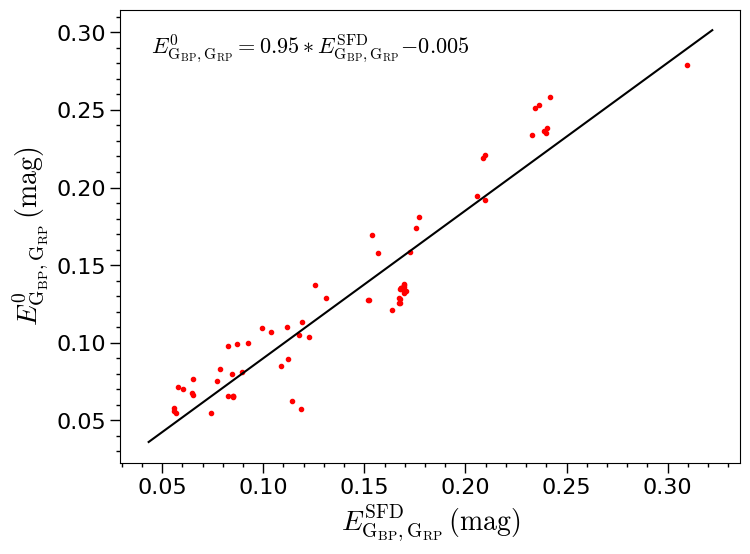}}
\caption{Comparison between $E_{\rm G_{BP},G_{RP}}^{\rm 0}$ and $E_{\rm G_{BP},G_{RP}}^{\rm SFD}$ in the $background\,\, region$.
\label{fig12}}
\end{figure}

The resultant $E_{\rm G_{BP},G_{RP}}^{\rm 0}$ is compared with \citet[SFD98]{1998ApJ...500..525S} for the $reference\,\,region$ since $E_{\rm G_{BP},G_{RP}}^{\rm 0}$ is supposed to be the cumulative extinction along the sightline. The value of $E_{\rm B,V}^{\rm SFD}$ is firstly converted to $E_{\rm G_{BP},G_{RP}}^{\rm SFD}$ for comparison \citep{2021ApJ...908L..14N}. Figure~\ref{fig12} shows the linear fitting between $E_{\rm G_{BP},G_{RP}}^{\rm 0}$ and the median value of $E_{\rm G_{BP},G_{RP}}^{\rm SFD}$ in each studied reference area. The slope is 0.95 and the intercept is -0.005, which means our result is very consistent with SFD98.

\subsection{Dust property of high-latitude molecular clouds}

\begin{figure}
\centering
\centerline{\includegraphics[scale=0.6]{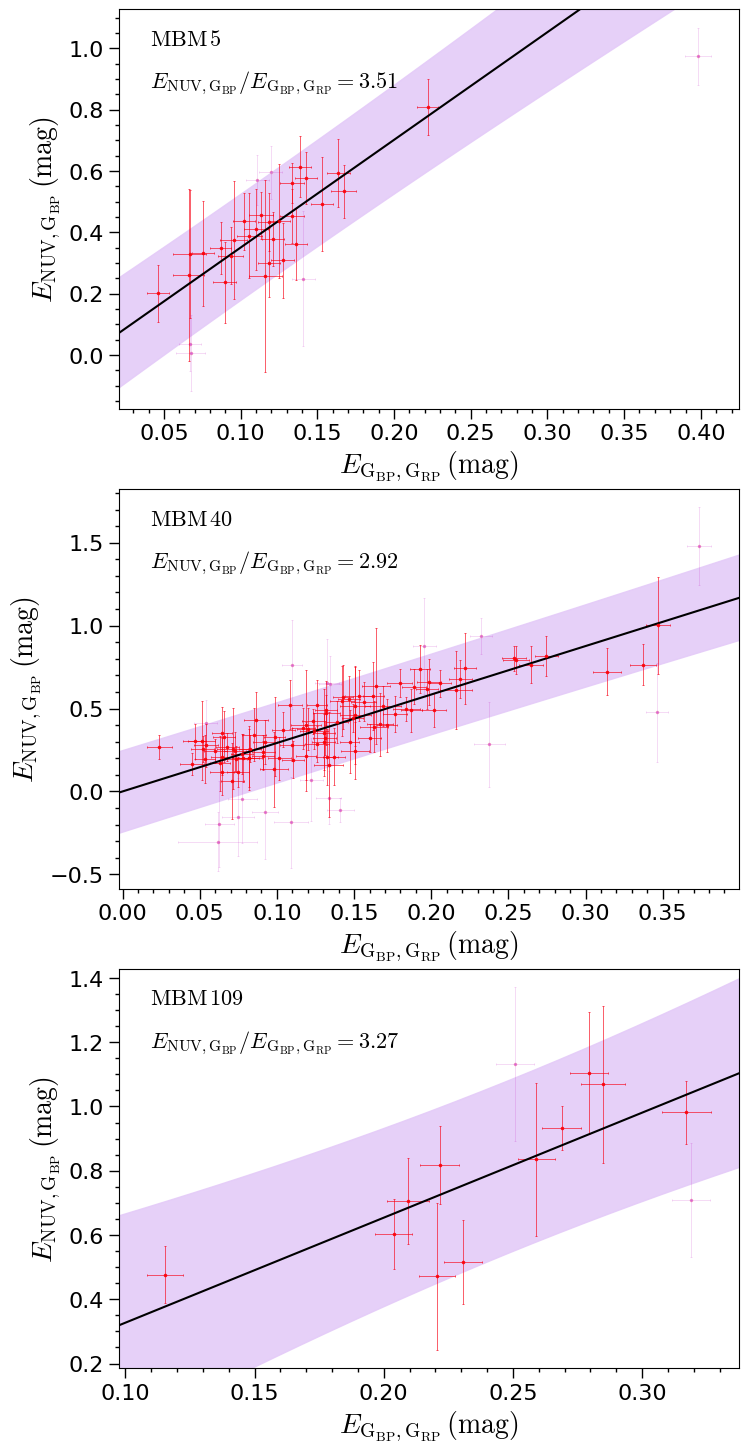}}
\caption{Linear fitting of the color excess $E_{{\rm NUV,G_{BP}}}$ to $ E_{{\rm G_{BP},G_{RP}}}$  in MBM5, MBM40 and MBM109. Red and magenta dots are sources in and out of the 95\% confidence interval, black line is the fitting curve, and the purple shadow is the 95\% confidence interval.
\label{fig13}}
\end{figure}

Because the extinction of a molecular cloud changes across the cloud,  the overall average color excess of the cloud, i.e. $\Delta E^{\rm MC}_{\rm G_{BP},G_{RP}}$ and $\Delta E^{\rm MC}_{\rm NUV,G_{BP}}$ should be an extinction indicator of the overall cloud. However, the ratio  $\Delta E^{\rm MC}_{\rm NUV,G_{BP}}/\Delta E^{\rm MC}_{\rm G_{BP},G_{RP}}$ of the cloud has great uncertainty perhaps due to very large dispersion in the extinction. Instead, the sources \emph{behind the cloud} are all taken into use to determine the color excess ratio $E_{\rm NUV,G_{BP}}$/$E_{\rm G_{BP},G_{RP}}$ and then the dust property of a molecular cloud. By subtracting the color excess of the diffuse interstellar medium at equal distances from the color excess of these sources, the corresponding color excess of the molecular cloud in the sightline is obtained. Requiring the number of selected sources N$\geq$3, the color excess ratios of 39 molecular clouds are calculated by linear fitting through iterative 3$\sigma$ clipping (\citetalias{2021ApJS..254...38S}). The results of MBM5, MBM40 and MBM109 are displayed in Figure~\ref{fig13}, where $E_{\rm NUV,G_{BP}}$ and $E_{\rm G_{BP},G_{RP}}$ have a good linear relationship.

\begin{figure}
\centering
\centerline{\includegraphics[scale=0.85]{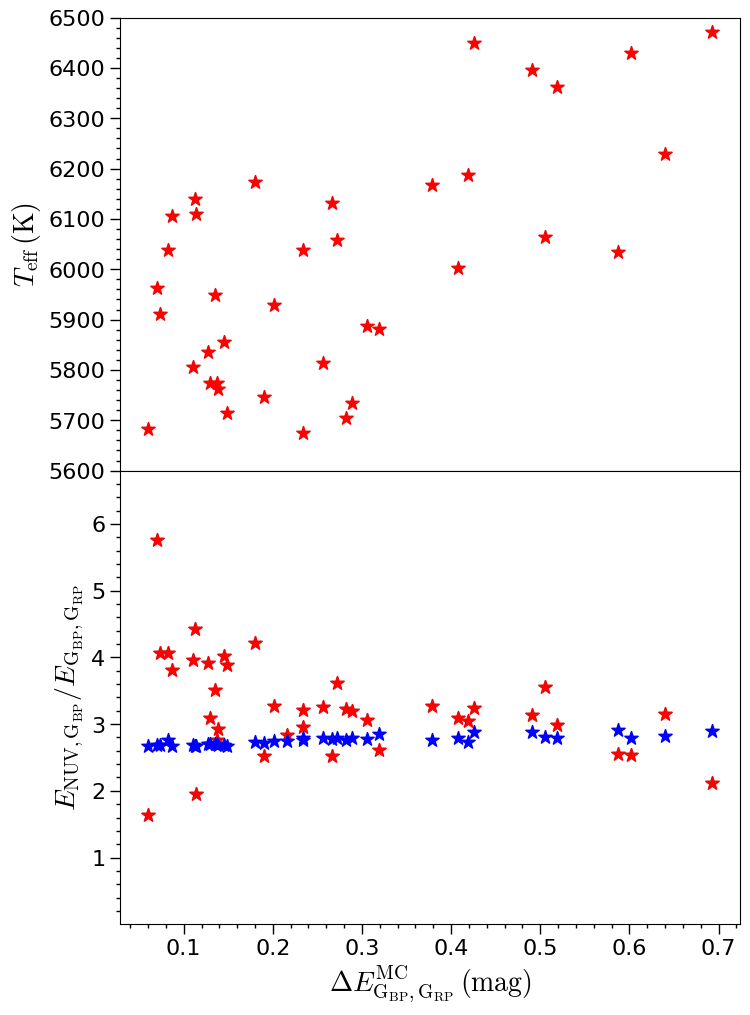}}
\caption{The change of $\Teff$ (upper panel) and color excess ratios $E_{\rm NUV,G_{BP}}$/$E_{\rm G_{BP},G_{RP}}$ (lower panel) with $\Delta E^{\rm MC}_{\rm G_{BP},G_{RP}}$. Red asterisks are the results of the studied molecular clouds, while the blue asterisks are the simulation results considering the effective wavelength shift because of $\Teff$ and $\Delta E^{\rm MC}_{\rm G_{BP},G_{RP}}$  (see \citetalias{2021ApJS..254...38S} for details).
\label{fig14}}
\end{figure}

The change of color excess ratios ($E_{\rm NUV,G_{BP}}/E_{\rm G_{BP},G_{RP}}$) with $\Delta E^{\rm MC}_{\rm G_{BP},G_{RP}}$  of these molecular clouds are shown in the lower panel of Figure~\ref{fig14}. Obviously, the color excess ratio of the molecular clouds (red asterisks) increases at $\Delta E^{\rm MC}_{\rm G_{BP},G_{RP}}< 0.3$. As \citetalias{2021ApJS..254...38S} pointed out, there exists systematic variation in both $\Teff$ and $E_{\rm G_{BP},G_{RP}}$ with the Galactic location for the tracing stars which can shift the effective wavelength of the filters and then the color excess ratio $E_{\rm NUV,G_{BP}}/E_{\rm G_{BP},G_{RP}}$ in a complicated way (see Figure 10 of \citetalias{2021ApJS..254...38S}). In brief, $E_{\rm NUV,G_{BP}}/E_{\rm G_{BP},G_{RP}}$ generally decreases with $E_{\rm G_{BP},G_{RP}}$ when $\Teff > 6500$ \,K, while increases with $E_{\rm G_{BP},G_{RP}}$ when $\Teff < 6500$\,K. On average, this color excess ratio is bigger for higher $\Teff$. The upper panel of  Figure~\ref{fig14} confirms that $\Teff$ changes with $\Delta E^{\rm MC}_{\rm G_{BP},G_{RP}}$.  In order to see the change of dust property, the effects by $\Teff$ and $\Delta E^{\rm MC}_{\rm G_{BP},G_{RP}}$ on the color excess ratio should be stripped off in advance. For this purpose, the color excess ratio of each is calculated assuming that only $\Teff$ and $\Delta E^{\rm MC}_{\rm G_{BP},G_{RP}}$ play a role, i.e. convolving the stellar emergent spectrum with the response curve of the filter and the F99 extinction curve at $R_{\rm V}=3.1$ \citep{1999PASP..111...63F}  to get the color excess ratio at corresponding effective wavelength. The derived color excess ratio is represented by blue asterisks in Figure~\ref{fig14}, which agrees with the expectation from \citetalias{2021ApJS..254...38S} that the lower $\Teff$ around 6000\,K in combination with smaller $\Delta E^{\rm MC}_{\rm G_{BP},G_{RP}}$ should lead to smaller $E_{\rm NUV,G_{BP}}/E_{\rm G_{BP},G_{RP}}$. The observed trend that the color excess ratio of the molecular clouds increases at smaller $E_{\rm G_{BP},G_{RP}}$ is thus on the opposite direction. It seems this trend can only be explained by the change of dust property at small extinction. In principle, the color excess ratio is sensitive to the composition and size distribution of the dust particles \citep{2011piim.book.....D}. \citet{2005ApJ...623..897L} conclude that many high-latitude clouds have enhanced abundances of relatively small grains based on the near-infrared extinction curves. The enhanced proportion of small grains can also explain the observed ratio variation of the near-ultraviolet-to-visual extinction here. Indeed, this trend is already found in the whole high-latitude sky in \citetalias{2021ApJS..254...38S} and now confirmed by this work.

\section{Summary}

This work uses the color excesses of more than four million stars in visual and one million stars in ultraviolet to explore the high-latitude molecular clouds cataloged by \citet{1985ApJ...295..402M}. The cloud and reference region are selected from the Planck/875GHz image in order to clarify the extinction caused by the cloud.   The distances to 66 clouds are determined by the extinction jump along the sightline caused by the cloud denser than the diffuse area.

The major results of this paper are as follows:

\begin{enumerate}

\item The cumulative color excess $E_{\rm G_{BP},G_{RP}}^{\rm 0}$ of the diffuse ISM and scale hight $h_{visible}$ of dust disk is derived for 66 areas, while the cumulative color excess $E_{\rm NUV,G_{BP}}^{\rm 0}$ of the diffuse ISM is obtained for 39 areas. The calculated scale height is  around 50-250 pc that agrees with the thin gaseous disk of the Milky.

\item The distances and color excess $\Delta E^{\rm MC}_{\rm G_{BP},G_{RP}}$ is determined for 66 molecular clouds, and the extinction jump $\Delta E^{\rm MC}_{\rm NUV,G_{BP}}$ is determined for 39 molecular clouds. The distances of this work are slightly larger than the results of \citet{2014ApJ...786...29S} and closer to that of \citet{2019ApJ...879..125Z}.

\item The color excess ratio $E_{\rm G_{BP},G_{RP}}$/$E_{\rm NUV,G_{BP}}$ of 39 molecular clouds is calculated and found to be obviously larger at lower extinction, which can not be interpreted by the shift of effective wavelength because of the variation in $\Teff$ and $E_{\rm G_{BP},G_{RP}}$. This indicates that the molecular clouds with lower extinction has more small dust particles.

\end{enumerate}


\acknowledgments{We thank Profs. Jian Gao and Haibo Yuan, and Mr. Ye Wang and Bin Yu for their discussions. We are also grateful to the referee for his/her helpful suggestions. 
This work is supported by CMS-CSST-2021-A09, National Key R\&D Program of China No. 2019YFA0405503  and NSFC 11533002. HZ is funded by the China Scholarship Council (No. 201806040200). This work made use of the data taken by \emph{GALEX}, LAMOST, \emph{Gaia} and GALAH, MBM.
}


\facilities{\emph{GALEX}, LAMOST, \emph{Gaia}, GALAH, MBM}

\clearpage
\bibliographystyle{aasjournal}
\bibliography{ccm}


\begin{deluxetable}{llllllllllllllllllll}
\rotate 
\caption{\label{table1}The distances and color excesses of the molecular clouds}
\tablehead{\colhead{Cloud{\tablenotemark{a}}} & \colhead{$l${\tablenotemark{a}}} & \colhead{$b${\tablenotemark{a}}}  & \colhead{$d_{Zucker}${\tablenotemark{b}}}  &  \colhead{$d_{Schlafly}${\tablenotemark{c}}}  & \colhead{$d_{this\,\, work}${\tablenotemark{d}}}& \colhead{$E_{\rm G_{BP},G_{RP}}^{\rm 0}${\tablenotemark{d}}}  & \colhead{$h'_{\rm G_{BP},G_{RP}}${\tablenotemark{d}}}  & \colhead{$\Delta E^{\rm MC}_{\rm G_{BP},G_{RP}}${\tablenotemark{d}}}  & \colhead{m{\tablenotemark{d}}}  & \colhead{$E_{\rm NUV,G_{BP}}^{\rm 0}${\tablenotemark{d}}} & \colhead{$h'_{\rm NUV,G_{BP}}${\tablenotemark{d}}}  & \colhead{$\Delta E^{\rm MC}_{\rm NUV,G_{BP}}${\tablenotemark{d}}}  & \colhead{CERs{\tablenotemark{d}}} \\
		\colhead{ } & \colhead{$(^{\circ})$}& \colhead{$(^{\circ})$}& \colhead{$(\rm pc)$} & \colhead{$(\rm pc)$} & \colhead{$(\rm pc)$} & \colhead{$(\rm mag)$} & \colhead{$(\rm pc)$} & \colhead{$(\rm mag)$} & \colhead{ } & \colhead{$(\rm mag)$} & \colhead{$(\rm pc)$} & \colhead{$(\rm mag)$} & \colhead{ }\\
		\colhead{(1)} & \colhead{(2)}& \colhead{(3)} & \colhead{(4)} & \colhead{(5)}  & \colhead{(6)} & \colhead{(7)}  & \colhead{(8)}  & \colhead{(9)}
& \colhead{(10)}  & \colhead{(11)}  & \colhead{(12)} & \colhead{(13)}
}
\startdata
${\rm MBM}  1$ & 110.19 & -41.229 & $265$ & $228$ & $285\pm9.0$ & $0.07\pm0.0005$ & $146\pm8.7$ & $0.09\pm0.0022$ & 8.0 & $0.20\pm0.0155$ & $296\pm81.3$ & $0.32\pm0.0278$ & $3.81$ \\
${\rm MBM}  3$ & 131.291 & -45.676 & $314$ & $277$ & $309\pm1.4$ & $0.07\pm0.0003$ & $193\pm3.8$ & $0.11\pm0.0010$ & 4.0 & $0.20\pm0.0036$ & $185\pm16.2$ & $0.40\pm0.0082$ & $3.96$ \\
${\rm MBM}  4$ & 133.515 & -45.303 & $286$ & $269$ & $320\pm8.0$ & $0.08\pm0.0014$ & $274\pm14.6$ & $0.13\pm0.0013$ & 4.0 & $0.22\pm0.0133$ & $104\pm44.2$ & $0.44\pm0.0183$ & $3.92$ \\
${\rm MBM}  5$ & 145.967 & -49.074 & $279$ & $187$ & $297\pm2.3$ & $0.08\pm0.0003$ & $225\pm3.1$ & $0.14\pm0.0012$ & 4.0 & $0.23\pm0.0051$ & $260\pm19.7$ & $0.46\pm0.0188$ & $3.51$ \\
${\rm MBM}  6$ & 145.065 & -39.349 & $111$ & $151$ & $153\pm0.9$ & $0.10\pm0.0003$ & $164\pm2.7$ & $0.23\pm0.0017$ & 8.0 & $0.28\pm0.0054$ & $179\pm19.2$ & $0.63\pm0.0251$ & $2.96$ \\
${\rm MBM}  7$ & 150.429 & -38.074 & $171$ & $148$ & $213\pm26.2$ & $0.11\pm0.0003$ & $183\pm2.1$ & $0.20\pm0.0026$ & 10.0 & $0.31\pm0.0039$ & $156\pm12.1$ & $0.63\pm0.0188$ & $3.28$ \\
${\rm MBM}  8$ & 151.75 & -38.669 & $255$ & $199$ & $262\pm0.2$ & $0.17\pm0.0007$ & $176\pm3.5$ & $0.26\pm0.0028$ & 12.0 & $0.48\pm0.0106$ & $108\pm22.6$ & $0.76\pm0.0742$ & $3.25$ \\
${\rm MBM}  9$ & 156.531 & -44.722 & $262$ & $246$ & $248\pm36.1$ & $0.14\pm0.0011$ & $223\pm7.7$ & $0.08\pm0.0034$ & 4.0 & $0.38\pm0.0282$ & $277\pm67.3$ & $0.35\pm0.0440$ & $4.07$ \\
${\rm MBM} 11$ & 157.983 & -35.06 & $250$ & $185$ & $147\pm101.7$ & $0.11\pm0.0016$ & $198\pm14.2$ & $0.33\pm0.0037$ & 16.0 &  &  &  &  \\
${\rm MBM} 12$ & 159.351 & -34.324 & $252$ & $234$ & $278\pm61.8$ & $0.16\pm0.0008$ & $155\pm5.3$ & $0.51\pm0.0181$ & 4.0 & $0.49\pm0.0123$ & $201\pm23.5$ & $0.99\pm0.0451$ & $3.56$ \\
${\rm MBM} 13$ & 161.591 & -35.89 & $237$ & $191$ & $409\pm0.5$ & $0.17\pm0.0011$ & $177\pm4.7$ & $0.42\pm0.0108$ & 12.0 & $0.48\pm0.0162$ & $155\pm23.3$ & $0.70\pm0.0631$ & $3.24$ \\
${\rm MBM} 14$ & 162.458 & -31.861 & $275$ & $233$ & $295\pm0.4$ & $0.22\pm0.0009$ & $212\pm2.8$ & $0.27\pm0.0006$ & 2.0 & $0.69\pm0.0058$ & $259\pm7.6$ & $0.99\pm0.0168$ & $3.61$ \\
${\rm MBM} 15$ & 191.666 & -52.294 & $200$ & $160$ & $164\pm120.7$ & $0.07\pm0.0009$ & $191\pm10.2$ & $0.11\pm0.0038$ & 5.0 & $0.17\pm0.0136$ & $152\pm72.3$ & $0.22\pm0.0495$ & $1.96$ \\
${\rm MBM} 16$ & 170.603 & -37.273 & $170$ & $147$ & $210\pm28.4$ & $0.16\pm0.0004$ & $233\pm2.5$ & $0.64\pm0.0057$ & 10.0 & $0.47\pm0.0075$ & $246\pm15.8$ & $1.45\pm0.0442$ & $3.15$ \\
${\rm MBM} 17$ & 167.526 & -26.606 & $130$ & $165$ & $231\pm38.6$ & $0.22\pm0.0012$ & $260\pm5.4$ & $0.30\pm0.0072$ & 4.0 &  &  &  &  \\
${\rm MBM} 18$ & 189.105 & -36.016 & $155$ & $166$ & $149\pm1.9$ & $0.08\pm0.0007$ & $333\pm8.4$ & $0.41\pm0.0022$ & 12.0 & $0.21\pm0.0056$ & $336\pm28.4$ & $1.10\pm0.0148$ & $3.09$ \\
${\rm MBM} 19$ & 186.041 & -29.929 & $143$ & $156$ & $293\pm0.2$ & $0.08\pm0.0008$ & $368\pm10.3$ & $0.34\pm0.0075$ & 72.0 &  &  &  &  \\
${\rm MBM} 22$ & 208.091 & -27.477 & $266$ & $238$ & $181\pm1.2$ & $0.06\pm0.0114$ & $388\pm137.2$ & $0.11\pm0.0047$ & 2.0 &  &  &  &  \\
${\rm MBM} 23$ & 171.835 & 26.706 & $349$ & $305$ & $252\pm182.1$ & $0.10\pm0.0008$ & $395\pm10.1$ & $0.07\pm0.0052$ & 10.0 & $0.31\pm0.0167$ & $470\pm70.3$ & $0.27\pm0.0619$ & $5.75$ \\
${\rm MBM} 24$ & 172.272 & 26.965 & $351$ & $279$ & $338\pm0.9$ & $0.10\pm0.0010$ & $368\pm13.0$ & $0.11\pm0.0015$ & 4.0 & $0.32\pm0.0224$ & $490\pm77.1$ & $0.43\pm0.0273$ & $4.42$ \\
${\rm MBM} 25$ & 173.752 & 31.475 & $342$ & $297$ & $362\pm2.1$ & $0.06\pm0.0007$ & $317\pm12.2$ & $0.07\pm0.0008$ & 4.0 & $0.15\pm0.0101$ & $410\pm83.5$ & $0.28\pm0.0106$ & $4.06$ \\
${\rm MBM} 34$ & 2.307 & 35.7 & $117$ & $110$ & $178\pm43.3$ & $0.06\pm0.0004$ & $107\pm8.2$ & $0.14\pm0.0022$ & 14.0 & $0.14\pm0.0092$ & $174\pm69.1$ & $0.37\pm0.0283$ & $2.76$ \\
${\rm MBM} 35$ & 6.571 & 38.128 & $ 86$ & $ 89$ & $296\pm10.4$ & $0.19\pm0.0016$ & $150\pm9.5$ & $0.23\pm0.0066$ & 4.0 &  &  &  &  \\
${\rm MBM} 36$ & 4.229 & 35.792 & $107$ & $105$ & $ 99\pm10.6$ & $0.10\pm0.0006$ & $176\pm5.8$ & $0.42\pm0.0013$ & 8.0 & $0.34\pm0.0139$ & $234\pm38.0$ & $0.94\pm0.0356$ & $3.05$ \\
${\rm MBM} 37$ & 6.067 & 36.757 & $115$ & $121$ & $143\pm0.4$ & $0.11\pm0.0013$ & $ 83\pm13.7$ & $0.32\pm0.0011$ & 4.0 & $0.34\pm0.0290$ & $107\pm41.5$ & $0.80\pm0.0649$ & $2.60$ \\
${\rm MBM} 38$ & 8.222 & 36.338 & $ 92$ & $ 77$ & $286\pm14.2$ & $0.13\pm0.0009$ & $142\pm7.6$ & $0.52\pm0.0399$ & 12.0 & $0.40\pm0.0186$ & $108\pm50.2$ & $1.51\pm0.0675$ & $2.98$ \\
${\rm MBM} 40$ & 37.57 & 44.667 & $ 93$ & $ 64$ & $ 63\pm51.3$ & $0.06\pm0.0003$ & $122\pm6.4$ & $0.14\pm0.0013$ & 8.0 & $0.17\pm0.0046$ & $217\pm26.1$ & $0.39\pm0.0123$ & $2.92$ \\
${\rm MBM} 49$ & 64.496 & -26.539 & $212$ & $204$ & $330\pm2.1$ & $0.09\pm0.0006$ & $172\pm10.0$ & $0.13\pm0.0014$ & 2.0 & $0.24\pm0.0112$ & $179\pm55.3$ & $0.29\pm0.0258$ & $3.09$ \\
${\rm MBM} 51$ & 73.313 & -51.526 &  &  & $190\pm9.2$ & $0.07\pm0.0007$ & $100\pm9.0$ & $0.06\pm0.1058$ & 12.0 & $0.25\pm0.0233$ & $272\pm94.6$ & $0.14\pm0.0825$ & $1.64$ \\
\enddata
\tablenotetext{a}{The molecular cloud's series number (Col. 1) and  Galactic coordinates (Cols. 2 and 3) retrieved from  \citet{1985ApJ...295..402M}.}
\tablenotetext{b}{The distance and the error (Col. 4)  from \citet{2019ApJ...879..125Z}.}
\tablenotetext{c}{The distance and the error (Col. 5)  from \citet{2014ApJ...786...29S}.}
\tablenotetext{d}{The distance and the errors (Col. 6), the foreground fitting parameters (Col. 7 and 8) and the color excess jump in the optical (Col. 9),  the times of cloud's angular diameter (Col. 10), the foreground fitting parameters (Col. 11 and 12) and the color excess jump (Col. 13) in the optical-ultraviolet bands, and the color excess ratio of molecular clouds (Col. 14) from this work.}
\end{deluxetable}

\begin{deluxetable}{llllllllllllllllllll}
\rotate
\tablehead{\colhead{Cloud{\tablenotemark{a}}} & \colhead{$l${\tablenotemark{a}}} & \colhead{$b${\tablenotemark{a}}}  & \colhead{$d_{Zucker}${\tablenotemark{b}}}  &  \colhead{$d_{Schlafly}${\tablenotemark{c}}}  & \colhead{$d_{this\,\, work}${\tablenotemark{d}}}& \colhead{$E_{\rm G_{BP},G_{RP}}^{\rm 0}${\tablenotemark{d}}}  & \colhead{$h'_{\rm G_{BP},G_{RP}}${\tablenotemark{d}}}  & \colhead{$\Delta E^{\rm MC}_{\rm G_{BP},G_{RP}}${\tablenotemark{d}}}  & \colhead{m{\tablenotemark{d}}}  & \colhead{$E_{\rm NUV,G_{BP}}^{\rm 0}${\tablenotemark{d}}} & \colhead{$h'_{\rm NUV,G_{BP}}${\tablenotemark{d}}}  & \colhead{$\Delta E^{\rm MC}_{\rm NUV,G_{BP}}${\tablenotemark{d}}}  & \colhead{CERs{\tablenotemark{d}}} \\
		\colhead{ } & \colhead{$(^{\circ})$}& \colhead{$(^{\circ})$}& \colhead{$(\rm pc)$} & \colhead{$(\rm pc)$} & \colhead{$(\rm pc)$} & \colhead{$(\rm mag)$} & \colhead{$(\rm pc)$} & \colhead{$(\rm mag)$} & \colhead{ } & \colhead{$(\rm mag)$} & \colhead{$(\rm pc)$} & \colhead{$(\rm mag)$} & \colhead{ }\\
		\colhead{(1)} & \colhead{(2)}& \colhead{(3)} & \colhead{(4)} & \colhead{(5)}  & \colhead{(6)} & \colhead{(7)}  & \colhead{(8)}  & \colhead{(9)}
& \colhead{(10)}  & \colhead{(11)}  & \colhead{(12)} & \colhead{(13)}
}
\startdata
${\rm MBM} 53$ & 93.965 & -34.058 & $259$ & $253$ & $266\pm0.7$ & $0.07\pm0.0003$ & $204\pm6.3$ & $0.18\pm0.0013$ & 8.0 & $0.20\pm0.0071$ & $285\pm36.9$ & $0.85\pm0.0154$ & $4.21$ \\
${\rm MBM} 54$ & 91.624 & -38.103 & $245$ & $231$ & $238\pm18.5$ & $0.06\pm0.0005$ & $159\pm7.2$ & $0.15\pm0.0028$ & 10.0 & $0.14\pm0.0044$ & $163\pm35.6$ & $0.53\pm0.0137$ & $4.03$ \\
${\rm MBM} 55$ & 89.19 & -40.936 & $245$ & $206$ & $266\pm1.5$ & $0.06\pm0.0004$ & $152\pm6.9$ & $0.15\pm0.0007$ & 4.0 & $0.15\pm0.0069$ & $244\pm54.2$ & $0.22\pm0.0207$ & $3.88$ \\
${\rm MBM} 56$ & 103.075 & -26.06 & $265$ & $227$ & $271\pm46.4$ & $0.11\pm0.0006$ & $173\pm6.5$ & $0.19\pm0.0024$ & 4.0 & $0.33\pm0.0098$ & $163\pm40.1$ & $0.58\pm0.0714$ & $2.52$ \\
${\rm MBM}101$ & 158.191 & -21.412 & $289$ & $283$ & $288\pm0.2$ & $0.26\pm0.0004$ & $194\pm1.8$ & $0.60\pm0.0012$ & 8.0 & $0.81\pm0.0043$ & $245\pm5.3$ & $1.36\pm0.0738$ & $2.54$ \\
${\rm MBM}102$ & 158.561 & -21.154 & $289$ & $275$ & $289\pm0.1$ & $0.25\pm0.0004$ & $201\pm1.6$ & $0.59\pm0.0010$ & 8.0 & $0.79\pm0.0040$ & $248\pm6.0$ & $1.19\pm0.0480$ & $2.56$ \\
${\rm MBM}103$ & 158.885 & -21.552 & $279$ & $269$ & $285\pm0.1$ & $0.25\pm0.0006$ & $196\pm3.3$ & $0.49\pm0.0009$ & 8.0 & $0.78\pm0.0041$ & $235\pm5.5$ & $1.18\pm0.0368$ & $3.13$ \\
${\rm MBM}104$ & 158.405 & -20.436 & $281$ & $262$ & $291\pm0.1$ & $0.28\pm0.0007$ & $194\pm3.4$ & $0.69\pm0.0007$ & 5.0 & $0.95\pm0.0110$ & $243\pm11.4$ & $1.09\pm0.0628$ & $2.11$ \\
${\rm MBM}105$ & 169.52 & -20.126 & $127$ & $139$ & $142\pm0.4$ & $0.20\pm0.0003$ & $178\pm1.6$ & $0.27\pm0.0005$ & 5.0 & $0.61\pm0.0073$ & $192\pm13.3$ & $0.62\pm0.0134$ & $2.52$ \\
${\rm MBM}106$ & 176.334 & -20.781 & $158$ & $190$ & $179\pm0.1$ & $0.23\pm0.0006$ & $213\pm2.4$ & $0.40\pm0.0005$ & 16.0 &  &  &  &  \\
${\rm MBM}107$ & 177.654 & -20.343 & $141$ & $197$ & $142\pm0.3$ & $0.24\pm0.0009$ & $213\pm3.5$ & $0.48\pm0.0007$ & 16.0 &  &  &  &  \\
${\rm MBM}108$ & 178.238 & -20.342 & $143$ & $168$ & $139\pm0.4$ & $0.24\pm0.0009$ & $213\pm2.8$ & $0.48\pm0.0014$ & 16.0 &  &  &  &  \\
${\rm MBM}109$ & 178.93 & -20.1 & $155$ & $160$ & $176\pm8.5$ & $0.23\pm0.0005$ & $209\pm2.1$ & $0.38\pm0.0012$ & 16.0 & $0.72\pm0.0038$ & $241\pm4.9$ & $0.82\pm0.0367$ & $3.27$ \\
${\rm MBM}110$ & 207.598 & -22.944 & $356$ & $313$ & $300\pm1.4$ & $0.13\pm0.0025$ & $464\pm19.5$ & $0.12\pm0.0009$ & 4.0 &  &  &  &  \\
${\rm MBM}111$ & 208.547 & -20.222 & $400$ & $366$ & $403\pm0.3$ & $0.13\pm0.0025$ & $468\pm21.2$ & $0.30\pm0.0011$ & 4.0 &  &  &  &  \\
${\rm MBM}115$ & 342.331 & 24.146 & $141$ & $137$ & $126\pm46.4$ & $0.13\pm0.0006$ & $ 52\pm2.4$ & $0.28\pm0.0015$ & 8.0 & $0.37\pm0.0143$ & $ 76\pm22.1$ & $0.79\pm0.0362$ & $3.23$ \\
${\rm MBM}116$ & 342.715 & 24.506 & $137$ & $134$ & $158\pm29.6$ & $0.14\pm0.0006$ & $ 51\pm1.6$ & $0.29\pm0.0015$ & 8.0 & $0.39\pm0.0134$ & $ 70\pm17.8$ & $0.83\pm0.0383$ & $3.20$ \\
${\rm MBM}117$ & 343.001 & 24.085 & $138$ & $140$ & $141\pm2.1$ & $0.13\pm0.0006$ & $ 51\pm1.3$ & $0.23\pm0.0014$ & 8.0 & $0.40\pm0.0164$ & $ 86\pm25.6$ & $1.01\pm0.0832$ & $3.21$ \\
${\rm MBM}118$ & 344.018 & 24.758 & $140$ & $ 56$ & $146\pm10.3$ & $0.13\pm0.0007$ & $ 51\pm1.9$ & $0.27\pm0.0021$ & 8.0 &  &  &  &  \\
${\rm MBM}119$ & 341.613 & 21.396 & $169$ & $150$ & $111\pm84.2$ & $0.12\pm0.0009$ & $ 51\pm1.0$ & $0.11\pm0.0024$ & 8.0 &  &  &  &  \\
${\rm MBM}120$ & 344.231 & 24.188 & $135$ & $ 59$ & $145\pm0.7$ & $0.12\pm0.0007$ & $ 52\pm2.1$ & $0.28\pm0.0022$ & 8.0 &  &  &  &  \\
${\rm MBM}123$ & 343.281 & 22.121 & $143$ & $101$ & $ 75\pm63.9$ & $0.13\pm0.0008$ & $ 51\pm1.0$ & $0.22\pm0.0035$ & 8.0 & $0.50\pm0.0592$ & $197\pm81.9$ & $0.24\pm0.0562$ & $2.83$ \\
${\rm MBM}124$ & 343.966 & 22.725 & $145$ & $ 89$ & $147\pm122.3$ & $0.13\pm0.0006$ & $ 51\pm0.9$ & $0.14\pm0.0098$ & 8.0 &  &  &  &  \\
${\rm MBM}125$ & 355.536 & 22.541 & $129$ & $115$ & $131\pm1.1$ & $0.14\pm0.0004$ & $ 51\pm1.6$ & $0.30\pm0.0026$ & 21.0 &  &  &  &  \\
${\rm MBM}127$ & 355.409 & 20.877 & $146$ & $147$ & $150\pm0.2$ & $0.14\pm0.0003$ & $ 51\pm1.3$ & $0.89\pm0.0070$ & 21.0 &  &  &  &  \\
${\rm MBM}128$ & 355.562 & 20.592 & $136$ & $134$ & $150\pm0.2$ & $0.14\pm0.0003$ & $ 51\pm1.4$ & $0.89\pm0.0067$ & 21.0 &  &  &  &  \\
${\rm MBM}129$ & 356.155 & 20.761 & $139$ & $141$ & $145\pm0.7$ & $0.14\pm0.0004$ & $ 51\pm47.5$ & $0.52\pm0.0033$ & 21.0 &  &  &  &  \\
${\rm MBM}130$ & 356.805 & 20.265 & $129$ & $109$ & $150\pm0.1$ & $0.14\pm0.0004$ & $ 51\pm1.3$ & $0.60\pm0.0042$ & 21.0 &  &  &  &  \\
${\rm MBM}131$ & 359.156 & 21.787 & $158$ & $106$ & $161\pm0.3$ & $0.14\pm0.0004$ & $ 51\pm1.2$ & $0.48\pm0.0027$ & 21.0 &  &  &  &  \\
${\rm MBM}133$ & 359.176 & 21.37 & $161$ & $ 98$ & $240\pm0.6$ & $0.14\pm0.0004$ & $ 51\pm1.2$ & $0.57\pm0.0052$ & 21.0 &  &  &  &  \\
${\rm MBM}134$ & 0.132 & 21.782 & $158$ & $121$ & $285\pm0.7$ & $0.07\pm0.0003$ & $ 73\pm8.1$ & $0.46\pm0.0062$ & 24.0 &  &  &  &  \\
${\rm MBM}136$ & 1.271 & 20.992 & $139$ & $120$ & $110\pm20.2$ & $0.09\pm0.0004$ & $101\pm4.6$ & $0.42\pm0.0027$ & 21.0 &  &  &  &  \\
${\rm MBM}145$ & 8.482 & 21.842 & $108$ & $152$ & $185\pm0.6$ & $0.07\pm0.0005$ & $ 71\pm12.6$ & $0.49\pm0.0024$ & 16.0 &  &  &  &  \\
${\rm MBM}146$ & 8.784 & 22.035 & $116$ & $179$ & $197\pm0.4$ & $0.07\pm0.0005$ & $ 63\pm8.6$ & $0.46\pm0.0058$ & 16.0 &  &  &  &  \\
${\rm MBM}148$ & 7.543 & 21.066 & $156$ & $116$ & $186\pm0.4$ & $0.07\pm0.0007$ & $ 80\pm14.4$ & $0.55\pm0.0023$ & 16.0 &  &  &  &  \\
${\rm MBM}151$ & 21.533 & 20.93 & $138$ & $122$ & $145\pm0.2$ & $0.18\pm0.0003$ & $100\pm1.9$ & $0.31\pm0.0006$ & 8.0 & $0.53\pm0.0042$ & $107\pm9.6$ & $0.83\pm0.0135$ & $3.07$ \\
${\rm MBM}152$ & 359.48 & -20.474 &  &  & $ 86\pm65.8$ & $0.10\pm0.0021$ & $134\pm16.0$ & $0.15\pm0.0024$ & 4.0 &  &  &  &
\enddata
\end{deluxetable}

\begin{deluxetable}{llllll}
\caption{\label{table2}MBM molecular clouds associated with Lynds dark cloud}
\tablehead{\colhead{MBM{\tablenotemark{a}}} & \colhead{LDN{\tablenotemark{a}}} & \colhead{$Area^{\rm MBM}${\tablenotemark{b}}} & \colhead{$Area^{\rm LDN}${\tablenotemark{b}}} & \colhead{$E_{\rm MBM}^{max}${\tablenotemark{c}}} & \colhead{$E_{\rm LDN}^{max}${\tablenotemark{c}}}\\
		\colhead{ } & \colhead{ } & \colhead{$\rm deg^2$}& \colhead{$\rm deg^2$}  & \colhead{$\rm mag$} & \colhead{$\rm mag$}
}
\startdata
${\rm MBM} 12$ & ${\rm LDN}1453$ & 1.767 & 0.066 & 1.84 & 1.528 \\
${\rm MBM} 12$ & ${\rm LDN}1454$ & 1.767 & 0.86 & 1.84 & 1.84 \\
${\rm MBM} 12$ & ${\rm LDN}1457$ & 1.767 & 0.262 & 1.84 & 1.786 \\
${\rm MBM} 12$ & ${\rm LDN}1458$ & 1.767 & 0.056 & 1.84 & 1.113 \\
${\rm MBM} 18$ & ${\rm LDN}1569$ & 2.836 & 0.631 & 1.009 & 0.855 \\
${\rm MBM} 36$ & ${\rm LDN}134$ & 1.767 & 0.22 & 1.183 & 1.109 \\
${\rm MBM} 37$ & ${\rm LDN}169$ & 1.227 & 0.86 & 1.612 & 1.612 \\
${\rm MBM} 37$ & ${\rm LDN}183$ & 1.227 & 0.24 & 1.612 & 1.612 \\
${\rm MBM}101$ & ${\rm LDN}1452$ & 1.767 & 1.66 & 1.945 & 1.945 \\
${\rm MBM}101$ & ${\rm LDN}1448$ & 1.767 & 0.053 & 1.945 & 1.296 \\
${\rm MBM}101$ & ${\rm LDN}1451$ & 1.767 & 0.14 & 1.945 & 1.569 \\
${\rm MBM}102$ & ${\rm LDN}1448$ & 1.767 & 0.053 & 1.945 & 1.296 \\
${\rm MBM}102$ & ${\rm LDN}1451$ & 1.767 & 0.14 & 1.945 & 1.569 \\
${\rm MBM}102$ & ${\rm LDN}1452$ & 1.767 & 1.66 & 1.945 & 1.945 \\
${\rm MBM}103$ & ${\rm LDN}1451$ & 1.767 & 0.14 & 1.885 & 1.569 \\
${\rm MBM}103$ & ${\rm LDN}1448$ & 1.767 & 0.053 & 1.885 & 1.296 \\
${\rm MBM}103$ & ${\rm LDN}1452$ & 1.767 & 1.66 & 1.885 & 1.945 \\
${\rm MBM}104$ & ${\rm LDN}1452$ & 1.767 & 1.66 & 2.387 & 1.945 \\
${\rm MBM}107$ & ${\rm LDN}1543$ & 1.767 & 0.09 & 1.732 & 1.732 \\
${\rm MBM}107$ & ${\rm LDN}1546$ & 1.767 & 0.37 & 1.732 & 1.842 \\
${\rm MBM}108$ & ${\rm LDN}1543$ & 1.767 & 0.09 & 2.018 & 1.732 \\
${\rm MBM}108$ & ${\rm LDN}1546$ & 1.767 & 0.37 & 2.018 & 1.842 \\
${\rm MBM}109$ & ${\rm LDN}1546$ & 1.767 & 0.37 & 2.018 & 1.842 \\
${\rm MBM}110$ & ${\rm LDN}1634$ & 1.767 & 0.492 & 0.773 & 0.466 \\
${\rm MBM}111$ & ${\rm LDN}1640$ & 1.767 & 0.018 & 1.379 & 0.005 \\
${\rm MBM}125$ & ${\rm LDN}1721$ & 1.767 & 0.287 & 0.769 & 0.424 \\
${\rm MBM}126$ & ${\rm LDN}1719$ & 1.767 & 0.61 & 1.218 & 1.218 \\
${\rm MBM}127$ & ${\rm LDN}1719$ & 1.767 & 0.61 & 1.218 & 1.218 \\
${\rm MBM}128$ & ${\rm LDN}1719$ & 1.767 & 0.61 & 1.218 & 1.218 \\
${\rm MBM}129$ & ${\rm LDN}1719$ & 1.767 & 0.61 & 1.218 & 1.218 \\
${\rm MBM}130$ & ${\rm LDN}1752$ & 1.767 & 1.78 & 0.979 & 1.421 \\
${\rm MBM}131$ & ${\rm LDN}1781$ & 1.767 & 1.19 & 0.778 & 0.778 \\
${\rm MBM}133$ & ${\rm LDN}1781$ & 1.767 & 1.19 & 0.778 & 0.778 \\
${\rm MBM}134$ & ${\rm LDN}1781$ & 1.767 & 1.19 & 0.559 & 0.778 \\
${\rm MBM}145$ & ${\rm LDN}234$ & 1.767 & 1.41 & 0.648 & 0.802 \\
${\rm MBM}148$ & ${\rm LND}234$ & 1.767 & 1.41 & 0.802 & 0.802
\enddata
\tablenotetext{a}{The molecular cloud's series number (Col. 1 and 2)  from \citet{1985ApJ...295..402M} (MBM) and \citet{1962ApJS....7....1L} (LDN) .}
\tablenotetext{b}{The cloud area (Col. 3 and 4) from MBM and LDN.}
\tablenotetext{c}{The maximum color excess $E_{\rm G_{BP},G_{RP}}$ in cloud region (Col. 5 and 6)  from MBM and LDN.}
\end{deluxetable}

\clearpage

\clearpage
\end{CJK*}
\end{document}